\documentclass[11pt,english]{article}
\usepackage{amsmath}
\usepackage{amssymb}
\usepackage{amsthm}
\usepackage{graphicx}
\usepackage{mathrsfs}
\usepackage{diagxy}
\usepackage[all]{xy}
\usepackage[hang,stable]{footmisc}
\ifx\href\undefined\else\hypersetup{linktocpage=true}\fi

\theoremstyle{definition}
\newtheorem{theorem}{Theorem}
\newtheorem{definition}[theorem]{Definition}
\newtheorem{remark}[theorem]{Remark}

\newcommand{\Mink}{\M}
\newcommand{\EMT}{\mathbf{T}}
\newcommand{\scri}{\mathscr{I}}
\newcommand{\supp}{\mathrm{supp}}
\newcommand{\M}{\mathrm{M}}
\newcommand{\vecM}{\mathrm{V}}
\newcommand{\Lin}{\mathrm{Lin}}
\newcommand{\reals}{\mathbb{R}}
\newcommand{\integers}{\mathbb{Z}}
\newcommand{\rationals}{\mathbb{Q}}
\newcommand{\Frames}{\mathcal{F}}
\newcommand{\End}{\mathrm{End}}
\newcommand{\GL}{\mathrm{GL}}
\newcommand{\Aff}{\mathrm{Aff}}
\newcommand{\Trans}{\mathrm{Trans}}
\newcommand{\Aut}{\mathrm{Aut}}
\newcommand{\Lie}{\mathrm{Lie}}
\newcommand{\Lor}{\mathrm{Lor}}
\newcommand{\LieLor}{\mathfrak{lor}}
\newcommand{\Poin}{\mathrm{Poin}}
\newcommand{\LiePoin}{\mathfrak{poin}}
\newcommand{\MM}{\mathfrak{M}\,}% Momentum Map
\newcommand{\id}{\mathrm{id}}
\newcommand{\Bij}{\mathrm{Bij}}
\newcommand{\Stab}{\mathrm{Stab}}
\newcommand{\Orb}{\mathrm{Orb}}
\newcommand{\Fl}{\mathrm{Fl}}
\newcommand{\Ad}{\mathrm{Ad}}
\newcommand{\Vect}{\mathrm{Vec}}
\newcommand{\ad}{\mathrm{ad}}
\newcommand{\measure}{\mathrm{d}}
\newcommand{\trace}{\mathrm{Tr}}
\newcommand{\kernel}{\mathrm{Ker}}

\newcommand{\EMTForm}{\mathcal{T}}
\newcommand{\etaup}{\eta_{\scriptscriptstyle\uparrow}}
\newcommand{\etadown}{\eta_{\scriptscriptstyle\downarrow}}
\newcommand{\Alt}{\mathrm{Alt}}

\begin{document}
\title{
Energy-Momentum Tensors and Motion\\
in Special Relativity}

\author{Domenico Giulini\footnote{Email: giulini@itp.uni-hannover.de}\\
        Institute for Theoretical Physics\\
        Riemann Center for Geometry and Physics\\
        Leibniz University Hannover\\ 
        Appelstrasse 2, D-30167 Hannover, Germany\\
        and\\
        Center for Applied Space Technology and Microgravity\\
        University of Bremen\\
        Am Fallturm, D-28359 Bremen, Germany}

\date{\today}

\maketitle

\begin{abstract}
\noindent
The notions of ``motion'' and ``conserved quantities'', if applied to 
extended objects, are already quite non-trivial in Special Relativity. 
This contribution is meant to remind us on all the relevant 
mathematical structures and constructions that underlie these 
concepts, which we will review in some detail. Next to the 
prerequisites from Special Relativity, like \emph{Minkowski 
space} and its \emph{automorphism group},  this will include the 
notion of a \emph{body} in Minkowski space, the \emph{momentum map}, 
a characterisation of the habitat of \emph{globally conserved 
quantities} associated with Poincar\'e symmetry -- so called 
\emph{Poincar\'e charges} --, the frame-dependent decomposition 
of global angular momentum into \emph{Spin} and an orbital part, 
and, last not least, the likewise frame-dependent notion of 
\emph{centre of mass} together with a geometric description 
of the \emph{M{\o}ller Radius}, of which we also list some typical 
values. Two Appendices present some mathematical background material 
on Hodge duality and group actions on manifolds. This is a 
contribution to the book: 
\emph{Equations of Motion in Relativistic Gravity}, edited by
Dirk P\"utzfeld and Claus L\"ammerzahl, to be published by 
Springer Verlag. 
\end{abstract}
\bigskip

\newpage

\setcounter{tocdepth}{1}
\tableofcontents

\newpage

\section{Introduction}
\label{sec:Introduction}
\begin{figure}[h!]
\centering\includegraphics[width=0.80\linewidth]%
{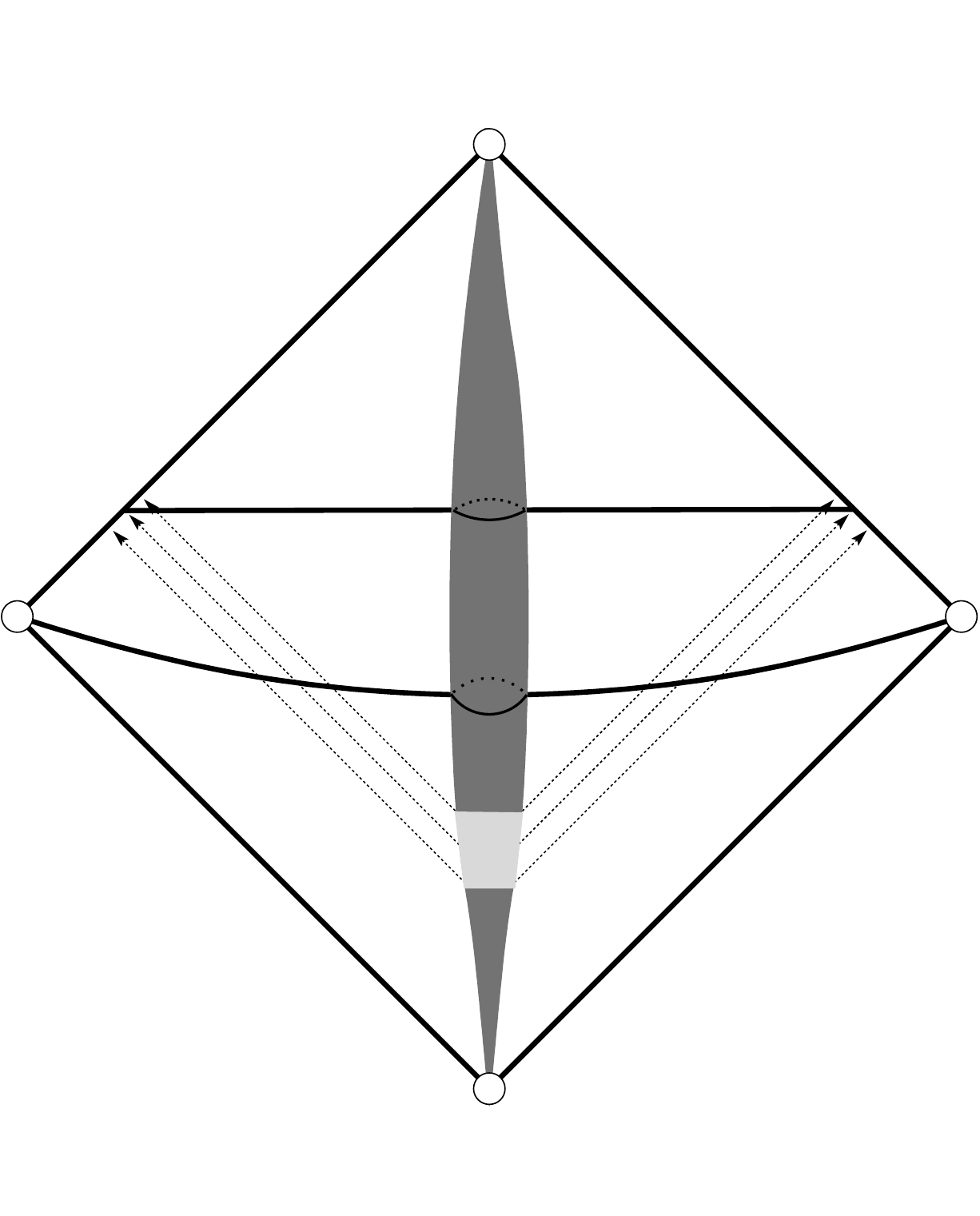}
\put(4,172){$I_0$}
\put(-302,172){$I_0$}
\put(-147,327){$I_+$}
\put(-147,15){$I_-$}
\put(-67,250){$\scri_+$}
\put(-233,250){$\scri_+$}
\put(-67,100){$\scri_-$}
\put(-233,100){$\scri_-$}
\put(-60,150){$S$}
\put(-240,150){$S$}
\put(-100,213){$L$}
\put(-200,213){$L$}
\caption{\label{fig:ConfMinkSpace_radiating}\small
History of a compact object in the conformal compactification 
of Minkowski space (Penrose Diagram). The five asymptotic 
regions of Minkowski space are future/past-timelike infinity 
$I_\pm$ (each a single point),  future/past-lightlike infinity 
$\scri_\pm$ (each a three-dimensional lightlike manifold of 
topology $\reals\times S^2$), and spacelike infinity $I_0$  
(a single point). The representation is not quite faithful 
because spacelike infinity, here represented by two points, 
is really just a single point. A faithful representation 
is obtained by wrapping the diamond-shaped 2-dimensional 
figure around a cylinder ($\reals\times S^1$), so as to 
identify both points $I_0$ of the diagram to a single one. 
$S$ and $L$ are both spacelike hypersurfaces stretching out 
to ``infinity''. But only $S$, which stretches out to spacelike 
infinity, is a Cauchy surface, i.e., covers all of spacetime 
in its domain of dependence.}  
\end{figure}
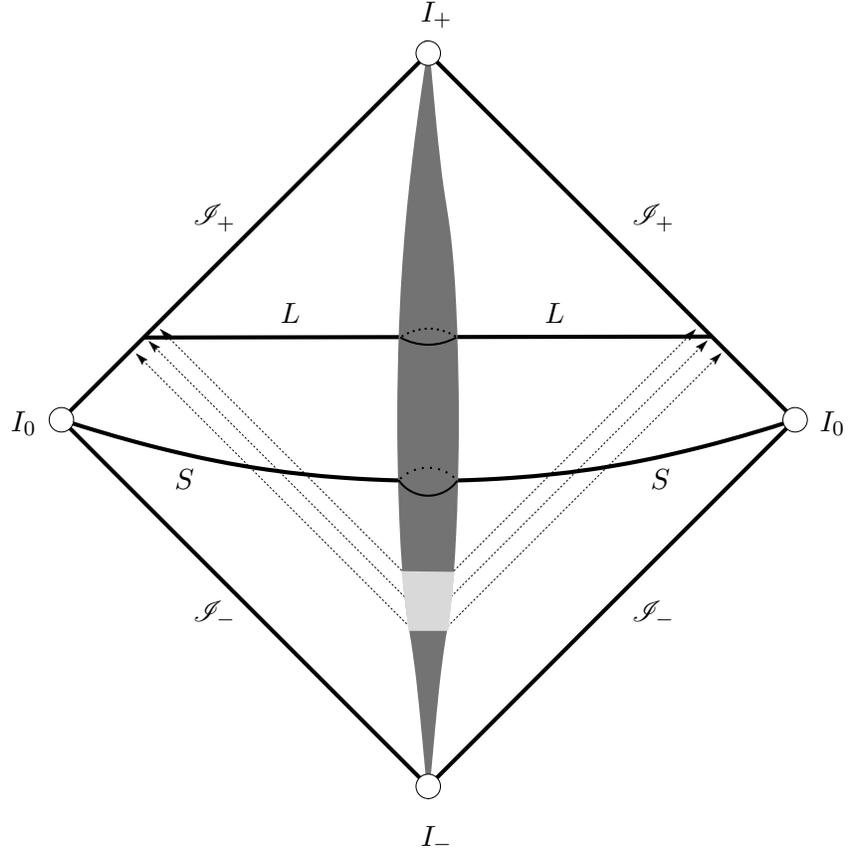

This contribution deals with the ``problem of motion'' in Special 
Relativity. Thus we work entirely in Minkowski space $\Mink$ (to 
be defined below) and represent a material system by an 
energy-momentum tensor $\EMT$ the support of which is to be 
identified with the set of events (points) in Minkowski space 
where matter ``exists'':
\begin{equation}
\label{eq:SupportEMT}
\supp(\EMT):=\overline{\{p\in\Mink \mid \EMT(p)\ne 0\}}\,.
\end{equation}    
A central assumption will be that the material system is 
\emph{spatially well localised}, which here shall mean 
that $\supp(\EMT)$ has compact intersection with any Cauchy 
hypersurface in $\Mink$. Note that Cauchy hypersurfaces 
end at spatial infinity $I_0$ and that $\supp(\EMT)$ need
not have compact intersection with asymptotically hyperboloidal 
spacelike hypersurfaces which tend to lightlike rather than 
spacelike infinity. This is depicted in 
Figure\,\ref{fig:ConfMinkSpace_radiating}.     
\begin{definition}
\label{def:Body}
We say that an energy-momentum tensor $\EMT$ describes a \emph{body}
iff the intersection of $\supp(\EMT)$ with any Cauchy hypersurface
in Minkowski space is compact.\qed
\end{definition} 
Hence we identify the event-set of a body with $\supp(\EMT)$,
which, in the sense made precise above, is of finite spatial 
extent, though it clearly will extend to timelike infinity. 
This is visualised as the a tubular neighbourhood stretching all 
the way from past-timelike to future-timelike infinity, as indicated 
by the shaded vertical tube in 
Figure\,\ref{fig:ConfMinkSpace_radiating}. 
It is also clear from Figure\,\ref{fig:ConfMinkSpace_radiating}
that we generally cannot require compact support of $\EMT$ on 
spacelike hypersurfaces which are not Cauchy, like $L$. In fact, if the body 
radiated in the finite past, given by the lighter-shaded part of the tubular 
region in the lower half of  Figure\,\ref{fig:ConfMinkSpace_radiating},
the radiation will propagate to $\scri_+$ and cover a neighbourhood 
in $L$ of its 2-sphere of intersection with  $\scri_+$, which is
of non-compact closure. This can be avoided for spacelike 
hypersurfaces ending at $I_0$ if we require a neighbourhood of 
$I_-$ to be free of radiation. This means that the body started 
to radiate a finite time in the past and that there is no incoming 
radiation from $\scri_-$ arbitrarily close to $I_0$. In fact, 
describing a quasi-isolated body would presumably mean to 
exclude incoming radiation altogether. This explains our 
motivation for Definition\,\ref{def:Body}. 

\newpage

A body should possess globally conserved quantities like 
linear and angular momentum. These are usually written 
down in a formulae like 
\begin{subequations}
\label{eq:NaiveCharges}
\begin{alignat}{1}
\label{eq:NaiveCharges-a}
P^a&\,=\,\int_\Sigma T^a_b\,u^b\,\measure\mu\,, \\
\label{eq:NaiveCharges-b}
J^{ab}[z]&\,=\,\int_\Sigma 
\bigl[(x^a-z^a)T^b_c-(x^b-z^b)T^a_c\bigr]u^c\,\measure\mu\,,
\end{alignat}
\end{subequations}
where $\Sigma$ is a Cauchy surface, $u^a$ are the components of its 
future-pointing normal, and $\measure\mu$ is the measure on $\Sigma$ 
induced from the ambient spacetime. See \cite{Dixon:SpecialRelativity} 
for a conceptually exceptionally clear discussion.

The problem with these expressions is that, on face value, they do 
not make any sense. For one thing, the integrands are vector/tensor 
valued, and adding them at different points does not result in 
anything with an obvious meaning. If we wish to interpret $P^a$
as the $a$-th (covariant) component of the vector of total linear 
momentum, we should characterise the vector space of which $P$ is 
an element. And, moreover, what does it mean to say that total linear 
(four-)momentum transforms like a four-vector (here covariant)?
Likewise, we wish to interpret $J^{ab}[z]$ as the $ab$-th
(contravariant) component of the antisymmetric 2nd-rank tensor of 
angular momentum with respect to the centre $z$. Again it is 
unclear what tensor space this $J[z]$ is an element of and what 
is meant by stating its representation property under 
Poincar\'e transformations. Are these spaces defined at points 
in spacetime, perhaps at ``infinity'', or in an abstract 
vector/tensor space globally associated to (but not \emph{in}) 
spacetime? Also, the difference $(x^a-z^a)$ that appears in  
 \eqref{eq:NaiveCharges-b} also makes no immediate sense. 
Is it supposed to be the $a$-th component of some 
``difference function'' on spacetime? Is it supposed to 
make sense in all coordinate systems, or just special ones;
and if the latter holds, what selects these special ones?

Clearly, all these questions do have answers, but these answers 
delicately depend on the precise mathematical structures with 
which spacetime is endowed. In our (highly idealised) case of  
Minkowski space, it is the high degree of symmetry of 
spacetime that allows us to naturally interpret 
\eqref{eq:NaiveCharges} so as to make unambiguous mathematical 
and physical sense. Removing or weakening these structures and 
pretending the expressions \eqref{eq:NaiveCharges} to still 
make sense without further qualifications means to commit a 
mathematical and conceptual sin. This does not mean that 
\eqref{eq:NaiveCharges} cannot be meaningfully generalised, 
but these generalisations will generally not be natural in a 
mathematical sense, that is, they will depend on additional 
structures and constructions to be imposed or selected 
``by hand''. The physical interpretation of what is then 
actually represented by the integrals  \eqref{eq:NaiveCharges} 
will delicately depend on these by-hand additions. It is 
therefore the aim of this introductory exposition to clarify 
the mathematical and physical meaning of \eqref{eq:NaiveCharges} 
in the simplest case, i.e. in Special Relativity. My strategy 
will be to fully display all the ingredients that go into the 
proper definition of 
\eqref{eq:NaiveCharges}. This, hopefully, will help to 
distinguish the generic difficulties of the gravitational 
case from those merely inherited from Special Relativity. 

\begin{figure}
\begin{center}
\includegraphics[height=0.4\linewidth]{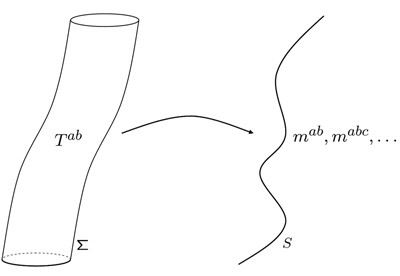}
\end{center}
\caption{As emphasised by this conference logo, a central problem
is to associate a timelike curve $S$ to the energy-momentum
tensor $T$. One would expect the line $S$ to lie in the 
``convex hull'' of the support of $T$, here represented by the
extended tube $\Sigma$.}
\label{nico:fig:2}
\end{figure}

Related to the issue of giving proper meaning to 
\eqref{eq:NaiveCharges} is the definition of ``centre of mass''
of an extended object. As you can see from its logo, this is a 
central concern of this conference (see Figure \ref{nico:fig:2}). 
If ``motion'' is the change of position in time, we need 
to be clear about how to define ``position'' in the first place. 
The issue of how to define position observables in any 
special-relativistic theory, classical and quantum, is 
notorious. See, e.g.,~\cite{Fleming:1965} for a good account.
In my contribution I will give a derivation of the M{\o}ller
radius which represents the ambiguity of defining position 
for systems with ``spin'', i.e., ``intrinsic angular momentum'', 
a notion also to be defined. So let us start at the beginning,
asking for the reader's patience!     

\newpage
 
\section{Minkowski space and Poincar\'e group}
In this section we wish to recall the definitions of Minkowski 
space and its automorphism group, despite the fact that this is 
generally considered a commonplace. But we think that there are 
some subtleties, in particular concerning the characterisation 
of its automorphism group, the \emph{Poincar\'e group}, that 
deserve to be said more than once. We start with 

\begin{definition}
\label{def:MinkowskiSpace}
\emph{Minkowski space} is a quadruple $(\M,\vecM,\eta,+)$, 
consisting of: 
\begin{enumerate}
\setlength{\itemsep}{-6pt}
\item
A set, $\M$, the elements of which are called spacetime 
points or events.
\item
A real 4-dimensional vector space $\vecM$.
\item
A simply transitive action of $\vecM$, considered as a 
group, on $\M$, denoted by $+$, i.e., 
\begin{equation}
\label{eq:VectorSpaceAction}
\M\times\vecM\rightarrow\M\,,\quad
(p,v)\mapsto p+v\,.
\end{equation}
\item
A non-degenerate symmetric bilinear form 
$\eta\in\vecM^*\otimes\vecM^*$ of signature $(+1,-1,-1,-1)$.\qed
\end{enumerate}
\end{definition}

\begin{remark}
\label{rem:MetricAndNotation}
Every non-degenerate bilinear form $\eta:V\times V\rightarrow\reals$
on a vector space $V$ defines an isomorphism $\etadown:V\rightarrow V^*$
to its dual space $V^*$ via the requirement $\etadown(v)(w):=\eta(v,w)$ 
for all $v,w\in V$; in short, $v\mapsto\etadown(v):=\eta(v,\cdot)$. 
Its inverse map is $\etaup: V^*\rightarrow V$, 
$\etaup:=\bigl(\etadown\bigr)^{-1}$, which in turn defines a 
non-degenerate bilinear form on the dual space, 
$\eta^{-1}:V^*\times V^*\rightarrow\reals$, via the requirement 
$\eta^{-1}(\alpha,\beta):=\alpha\bigl(\etaup(\beta)\bigr)$ for all 
$\alpha,\beta\in V^*$. On component-level this reads as follows: Let 
$\{e_a\mid 1\leq a\leq n\}$ be a basis of $V$ and $\{\theta^a\mid 1\leq a\leq n\}$ 
its dual basis of $V^*$, so that $\theta^a(e_b)=\delta^a_b$. Then, writing  
$v=v^ae_a$, we get  $\etadown(v)=v_b\theta^b$ with $v_b:=v^a\eta_{ab}$ 
and $\eta_{ab}:=\eta(e_a,e_b)$. Similarly, writing $\alpha=\alpha_a\theta^a$, 
we get $\etaup(\alpha)=\alpha^ae_a$
with $\alpha^a:=\eta^{ab}\alpha_b$ and $\eta^{ab}:=\eta^{-1}(\theta^a,\theta^b)$. 
This implies $\delta^a_b=\eta^{ac}\eta_{bc}=\eta^{ca}\eta_{cb}=\delta^a_b$
and, in particular, $\eta^{ab}=\eta^{ac}\eta^{bd}\eta_{cd}$
and $\eta_{ab}=\eta^{cd}\eta_{ca}\eta_{db}$. This explains why 
$\etaup$ and $\etadown$ are called the operations of ``index-raising'' 
and ``index lowering''. Sometimes the images of $\etaup$ and $\etadown$ 
are indicated by the musical symbols $\sharp$ (sharp) and $\flat$ (flat) 
respectively, i.e., one writes $\etaup(\alpha)=\alpha^\sharp$ and 
$\etadown(v)=v^\flat$, which makes sense as long as the bilinear form 
$\eta$ with respect to which these maps are defined is self understood.
We shall also employ this notation.  
Note that so far we did not assume $\eta$ to be symmetric, so that all 
formulae apply generally. However, from now on, and for the rest of this 
paper, the symbol $\eta$ shall always denote the Minkowski metric, which 
specialises the general case by symmetry and signature. Once $\eta$ is 
fixed, the isomorphisms between $V$ and $V^*$ as well as its extensions
to tensor products is clear from the context and it is sufficient and 
useful to use shorthand notatations, like 
$v\cdot w:=\eta(v,w)=v^\flat(w)$, $v^2:=v\cdot v$, and 
$\Vert v\Vert:=\sqrt{\vert v\cdot v\vert}$. Given 
$J=J^{ab}e_a\otimes e_b\in V\otimes V$ and $v\in V$, we shall also 
write $J\cdot v$ or $v\cdot J$ for the application of 
$J^{ab}e_a\otimes\etadown(e_b)=J^a_{\phantom{a} b}e_a\otimes\theta^b\in\End(V)$ or 
$J^{ab}e_b\otimes\etadown(e_a)=J_a^{\phantom{a} b}e_b\otimes\theta^a\in\End(V)$, 
respectively, to $v$. The inner products on $V$ and $V^*$ can be used to 
define inner products on any space built by taking tensor products 
of $V$ and $V^*$ just by slotwise contraction. However, in certain 
circumstances of high symmetry, e.g., for totally antisymmetric tensor 
products, it is more convenient to renormalise the slotwise inner 
product by combinatorial factors; like in formula \eqref{eq:RenormInnerProduct} of the Appendix.    
Finally we recall that the transposed of a general 
linear map $A:V\rightarrow W$ between vector spaces $V$ and $W$ is the 
linear map $A^\top:W^*\rightarrow V^*$, defined by 
$A^\top(\alpha):=\alpha\circ A$ for all $\alpha\in W^*$. There 
is a natural isomorphism between a vector space $V$ and its 
double dual $V^{**}$, so that 
we may identify these spaces without explicit mention. Symmetry of 
$\eta$ is then equivalent to $\etadown^\top=\etadown$ and symmetry of 
$\eta^{-1}$ to $\etaup^\top=\etaup$.

\end{remark}

\subsection{Affine spaces}
\label{sec:Affine-Spaces}
Note that 1.-3. define the notion of an affine space. Minkowski 
space is thus just a real 4-dimensional affine space, the associated 
vector space of which carries a Lorentz metric. Any vector space 
$\vecM$ is a group under addition, with group identity being 
given by the zero vector and the inverse of $v\in\vecM$ being 
$-v$. It is customary to use the same symbol, $+$, for the addition 
of vectors in $\vecM$ and the action of $\vecM$ on $\M$. 
This allows to write the action property in the intuitive form 
(compare Appendix\,\ref{sec:Appendix-GroupsActions} for the 
general definition of a group action on a set)
\begin{equation}
\label{eq:PlusNotation}
p+(v+w)=(p+v)+w=:p+v+w\,.\quad
\end{equation}  
But note the different meanings of $+$ in this equation. 
Moreover, we define the subtraction of a vector by the 
addition of the inverse:
\begin{equation}
\label{eq:MinusNotation-1}
p-v:=p+(-v)\,.
\end{equation}  
This allows one more simplifying notation: Since $\vecM$ acts 
simply transitive, there exists a \emph{unique} $v\in\vecM$ for any 
given pair $(p,q)\in\M\times\M$ so that $p=q+v$. We write 
\begin{equation}
\label{eq:MinusNotation-2}
v=p-q\,.
\end{equation}  
Hence the minus sign should be understood as 
\emph{difference map} $\M\times\M\rightarrow\vecM$, 
$(p,q)\rightarrow p-q$, defined through $p=q+(p-q)$.
Simple transitivity then implies 
\begin{equation}
\label{eq:MinusProperty-1}
(p-o)+(o-q)=p-q\,,
\end{equation}
which is equivalent to 
\begin{equation}
\label{eq:MinusProperty-2}
p+(q-o)=q+(p-o)\,.
\end{equation}
 
\subsection{Linear and affine frames}
\label{sec:Affine-Frames}
\begin{definition}
\label{def:AffineFrame}
A \emph{frame} $F$ for an affine space $(\M,\vecM,+)$ 
consists of a tuple $F=(o,f)$, where $o\in\M$ and 
$f\in\Lin(\reals^n,\vecM)$ is a frame of the vector space
$\vecM$. Recall that a frame $f$ of an $n$-dimensional real 
vector space $\vecM$ is an isomorphism from $\reals^n$ to 
$\vecM$. This is equivalent to choosing $n$ linear independent
vectors $\{e_1,\cdots,e_n\}\subset\vecM$, the images under $f$ 
of the canonical basis of $\reals^n$. The map $f$ is then 
defined by linear extension: $f(r^1,\cdots ,r^n)=\sum_{a=1}^nr^ne_n$. 
Its inverse map is given by $f^{-1}(v)=\bigl(\theta^1(v),\cdots,\theta^n(v)\bigr)$, where $\{\theta^1,\cdots, \theta^n\}\subset V^*$ is the dual basis of 
$\{e_1,\cdots,e_n\}$, i.e., $\theta^a(e_b)=\delta^a_b$. 
Similarly, an affine frame $F$ defines a bijective map between 
$\reals^n$ and the underlying set $\M$, denoted by the same 
letter $F$ and defined by 
\begin{equation}
\label{eq:DefAffineFrame}
\begin{split}
F:\reals^n\rightarrow\M\,,\quad
(r_1,\cdots,r_n)\mapsto F(r_1,\cdots,r_n)
:&=o+f(r^1,\cdots,r^n)\\
 &=o+\sum_{a=1}^{n}r^a e_a\,.
\end{split}
\end{equation}  
The inverse map is 
\begin{equation}
\label{eq:DefAffineFrame-Inverse}
\begin{split}
F^{-1}:\M\rightarrow\reals^n\,,\quad
p\mapsto F^{-1}(p)
:&= f^{-1}(p-o)\\
&=\bigl(\theta^1(p-o),\cdots,\theta^n(p-o)\bigr)\,.
\end{split}
\end{equation}  
Given two frames $F=(o,f)$ and $F'=(o',f')$, they are related by 
$F=F'\circ (F'^{-1}\circ F)$, where 
\begin{subequations}
\label{eq:DefAffineFrame-Relation}
\begin{equation}
\label{eq:DefAffineFrame-Relation-a}
F'^{-1}\circ F:\reals^n\rightarrow\reals^n\,,\quad
(r^¹,\cdots,r^n)\mapsto \bigl({r'}^1,\cdots,{r'}^n\bigr)
\end{equation}  
with 
\begin{equation}
\label{eq:DefAffineFrame-Relation-b}
\begin{split}
{r'}^a(r^1,\cdots,r^n)
&=[f'^{-1}(o-o')+f'^{-1}\circ f(r^1,\cdots,r^n)]^a\\
&=\theta'^a(o-o')+\sum_{b=1}^n {\theta'}^a(e_b)\,r^b\,.
\end{split}
\end{equation}
\end{subequations}
We denote the set of all affine frames of $\M$ by $\Frames_\M$.
\qed
\end{definition}  
\begin{remark}
\label{rem:TopologyDiffStructure}
Affine spaces naturally inherit a topology from $\reals^n$. 
It is defined to be the unique topology on $\M$ for which 
all frame maps \eqref{eq:DefAffineFrame} are homeomorphisms,
i.e., $F$ and $F^{-1}$ are continuous (hence $F$ is an open map). 
Note that if a particular $F'$ is a homeomorophism, than so 
is any other $F$, for  $F=F'\circ (F'^{-1}\circ F)$ and $F'^{-1}\circ F:\reals^n\rightarrow\reals^n$, 
given by \eqref{eq:DefAffineFrame-Relation-b}, is clearly a 
homeomorphism. Hence the open sets in $\M$ are precisely the 
images of open sets in $\reals^n$ under any $F$. Moreover, 
affine frames endow $\M$ with the structure of a smooth 
($C^\infty$, or even analytic) manifold since each frame 
defines a global chart with analytic transition functions  \eqref{eq:DefAffineFrame-Relation-b} between those charts.  
\qed
\end{remark}

\begin{definition}
\label{def:AffineCoordinates}
Affine frames define special, globally defined coordinates 
which are called \emph{affine coordinates} or, in a physical 
context, \emph{inertial coordinates}. Using these we may regard 
affine spaces as smooth ($C^\infty$, or even analytic) manifolds, 
as explained in Remark\,\ref{rem:TopologyDiffStructure}.
\end{definition}

Recall that the algebra of all linear self-maps of a vector 
space $\vecM$ onto itself is denoted by $\End(\vecM)$ 
(endomorphisms). The subset of all invertible elements in 
$\End(\vecM)$ is called $\GL(\vecM)$; it forms a group, 
the general-linear group (of self-isomorphisms, or 
Automorphisms regarding its structure as vector space) of 
$\vecM$. 
Accordingly, $\End(\reals^n)$ is just given by the algebra 
of all real $n\times n$ matrices and $\GL(\reals^n)$ by the 
group of all $n\times n$ matrices with non-vanishing determinant. 

A  frame of $\vecM$ defines an isomorphism of algebras 
$\End(\vecM)\rightarrow\End(\reals^n)$ through 
$A\mapsto A^f:=f\circ A\circ f^{-1}$. Its restriction to 
$\GL(\vecM)$ defines an isomorphism of groups 
$\GL(\vecM)\rightarrow\GL(\reals^n)$. Let us denote by 
$\Frames_\vecM$ the set of all frames of $\vecM$. There 
are two natural left actions of groups on $\Frames_\vecM$: 
$\GL(\vecM)$ acts on the left according to  
$(A,f)\mapsto A\circ f$ and $\GL(\reals^n)$ also acts 
on the left according to $(B,f)\mapsto f\circ B^{-1}$. 
Note that $(B,f)\mapsto f\circ B$ would be a right action; 
see Appendix\,\ref{sec:Appendix-GroupsActions} for a 
general discussions of group actions. 
Both actions commute and are each simply transitive. 
A combined left action of $\GL(\vecM)\times\GL(\reals^n)$ 
on  $\Frames_\vecM$ according to $\bigl((A,B),f\bigr)\mapsto
A\circ f\circ B^{-1}$ results. The action of $\GL(\reals^n)$ 
is sometimes called \emph{passive} since it merely 
moves the labels (coordinates) in label-space $\reals^n$, 
whereas $\GL(\vecM)$'s action is called \emph{active} since 
it really moves the points in the space $V$. Note 
that these adjectives refer to \emph{different} groups, 
which are isomorphic but not naturally so since picking 
any isomorphism requires extra choices to be made. 
For example, picking a frame $f$, an $f$-dependent isomorphism 
$\GL(\reals^n)\rightarrow\GL(\vecM)$ is defined through 
the stabiliser subgroup in  $\GL(\vecM)\times\GL(\reals^n)$
that fixes $f$ under the common left action just described. 
This isomorphism then simply reads 
$\GL(\reals^n)\ni B\mapsto A:=f\circ B\circ f^{-1}\in\GL(\vecM)$
(so that $A\circ f\circ B^{-1}=f$), which then also defines 
a frame-dependent left action of $GL(\reals^n)$ on $\vecM$.
With respect to the fixed frame $f$ the latter can then be 
used to define ``active'' transformations on $\vecM$ by means 
of what previously had been interpreted as mere label 
(coordinate) transformations. Failing to clearly state
the groups, their domains of action, and the structures 
to be considered fixed is often the source of considerable 
confusion regarding the distinction of ``active'' and 
``passive'' actions.

\subsection{Affine groups}
\label{sec:Affine-Group}
\begin{definition}
\label{def:AffineGroup}
Let $(\M,\vecM,+)$ be an $n$-dimensional real affine space. 
The \emph {affine  group}, denoted by $\Aff(\M)$, is the 
group of automorphisms of  $(\M,\vecM,+)$. This means that 
$\Aff(\M)$ is the subgroup of bijections of $\M$ preserving 
the simply transitive action $\vecM$ on $\M$. The word ``preserving''
means that for each $H\in\Aff(\M)$ there exists a unique 
$h\in\Aut(\vecM)$ so that $H(p+v)=H(p)+h(v)$ for all $p\in\M$ 
and all $v\in\vecM$. Here $\Aut(\vecM)$ is the 
automorphism group of $\vecM$, which is $\GL(\vecM)$ if we 
consider its structure as vector space or as topological 
group, i.e., $\GL(\vecM)$ are the continuous automorphisms of 
the topological group $V$. 
\begin{equation}
\label{eq:DefAffineMaps}
\begin{split}
\Aff(\M):=
\bigl\{H:\M\rightarrow\M \mid H(p+v)=H(p)+h(v)\,,h\in\GL(\vecM)\,,\forall v\in\vecM\bigr\}\,.
\end{split}
\end{equation}     
Note that this definition makes sense, for if $p'+v'=p+v$, or $(p'-p)+v'=v$, we have $H(p'+v')=H(p')+h(v')=H(p+(p'-p))+h(v')=H(p)+h((p'-p)+v')=H(p)+h(v)=H(p+v)$. 
\qed
\end{definition}

\begin{remark}
\label{rem:NonTrivialAutomorphisms}
We said that $\Aut(\vecM)$ is $\GL(\vecM)$ if we consider 
$\vecM$ either as vector space or as topological group, 
comprising all the continuous automorphisms in the latter case. 
This qualification is indeed necessary, for if we considered 
$\vecM$ merely as algebraic group, as it might seem sufficient 
at this point, $\Aut(\vecM)$ would indeed be very much larger 
than  $\GL(\vecM)$ in that it will also contain all the wildly 
discontinuous automorphisms that $V$ inherits from the likewise 
wildly discontinuous automorphisms of the algebraic group 
$(\reals,+)$. The latter are the discontinuous solutions 
$f:\reals\mapsto\reals$ to the so-called \emph{Cauchy functional equation}, 
$f(x+y)=f(x)+f(y)$, which are also bijections. 
It is elementary to show that all its solutions necessarily 
satisfy $f(qr)=qf(r)$ for all $q\in\rationals$ and all 
$r\in\reals$. This implies that $f(q)=qf(1)$, i.e., that $f$ is 
linear with slope $c:=f(1)$ on all rational numbers, and hence 
linear with slope $c$ on all real numbers if $f$ were required 
to be continuous (requiring continuity at one point is sufficient).  
Without requiring continuity we can only conclude that for 
fixed $r\in\reals$ and all $q\in\rationals$ we must have
$f(rq)=rq \bigl(f(r)/r\bigr)$, i.e., that $f$ is again linear 
on the $r$-multiples of the rationals, but now with possibly 
$r$-dependent slope $c(r):=f(r)/r$. Indeed, plenty of 
such discontinuous solutions exist and can be constructed as 
follows~\cite{Hamel:1905}: Consider $\reals$ as vector space 
over $\rationals$ and let $B\subset\reals$ be a (Hamel) 
basis, i.e., for each $r\in\reals$ there exists a unique 
\emph{finite} subset $\{e_1,\cdots, e_n\}\subset B$ and unique 
$(q_1,\cdots, q_n)\in\rationals^n$, such that $r=\sum_{i=1}^nq_ie_i$. 
As was shown in~\cite{Hamel:1905}, the existence of such a 
basis follows from the well-ordering theorem, though the 
cardinality of $B$ is that of $\reals$, i.e., the basis is 
uncountable. Now, any bijection $f:B\rightarrow B$ gives rise to 
an element of $\Aut(\reals,+)$ by uniquely extending $f$ from 
$B\subset\reals$ to $\reals$ in a $\rationals$-linear
fashion, i.e., by setting $f(\sum q_ie_i):=\sum q_if(e_i)$ for all 
finite linear combinations of elements in $B$ over $\rationals$. 
Moreover, if the initial permutation $f:B\rightarrow B$ is not 
linear, i.e., if the function $B\ni e\mapsto f(e)/e$ is not constant, 
the automorphism $f:\reals\rightarrow\reals$ so defined is ``wildly'' 
discontinuous, in the sense that its graph 
$\{(x,f(x))\mid x\in\reals\}\subset\reals^2$ is dense! In particular, 
given any $x\in\reals$, the image of \emph{any} intervall containing 
$x$ under $f$ is dense in $\reals$, no matter how small the intervall
was chosen to be. To see this, consider $e_1,e_2\in B$ so that 
$f(e_1)/e_1\ne f(e_2)/e_2$. Given any $(x,y)\in\reals^2$ we can 
uniquely solve the two equations $x=r_1e_1+r_2e_2$ and 
$y=r_1f(e_1)+r_2f(e_2)$, i.e., the single linear equation, 
\begin{equation}
\label{eq:DenseGraph}
\begin{pmatrix}
e_1&e_2\\
f(e_1)&f(e_2)
\end{pmatrix}
\begin{pmatrix}
r_1\\
r_2
\end{pmatrix}
=
\begin{pmatrix}
x\\
y
\end{pmatrix}\,,
\end{equation}
for $(r_1,r_2)\in\reals^2$ by rational operations, since the 
$2\times 2$ matrix in \eqref{eq:DenseGraph} is invertible. 
In particular, $(x,y)$ depends continuously on $(r_1,r_2)$ 
so that with rational $(q_1,q_2)\in\rationals^2$ in a 
neighbourhood of $(r_1,r_2)$ we get arbitrarily close to 
$(x,y)$, as was to be proven. All this implies that the usual 
abelian group structure underlying vector addition cannot be 
uniquely specified without requiring continuity. Interestingly 
this problem was first encountered in analytical mechanics 
in connection with attempts to mathematically characterise the 
law for the composition of forces~\cite{Darboux:1875} and only 
later recognised as essential for general axiomatic formulations 
of vector addition; see, e.g., \cite{Schimmack:1903}. 
For us all this means that we cannot avoid invoking a 
continuity hypothesis and that we must regard the abelian 
groups whose simply transitive action we require in the 
definition of affine spaces as topological groups acting 
continuously on affine space with its natural topology inherited 
from $\reals^n$; compare Remark\,\ref{rem:TopologyDiffStructure}.  
One might think that one gets away without continuity 
requirements if one defines $\Aff(\M)$ as that subgroup 
of the group of bijections (no continuity required here) of 
$\M$ which maps straight lines (physically: inertial trajectories) 
into straight lines (collinear sets of points into collinear sets
would also suffice). A classic result in affine geometry then 
tells us that such transformations necessarily coincide 
with the standard continuous affine transformations; see,
e.g., \cite{Berger:Geometry1}. However, here a continuity 
requirement has tacitly slipped into the notion of ``straight line'' 
(inertial trajectory), which in affine space is defined to 
be the orbit of a \emph{continuous} one-parameter subgroup of 
$\vecM$.         
\qed
\end{remark}

Coming back to the group of affine automorphisms as defined above. 
we see that Hence an element $H\in\Aff(\M)$ is uniquely 
specified by an ordered pair of points $(p,q)\in\M\times\M$ and 
an element $h\in\GL(\vecM)$. The second point $q$ is regarded as 
the image of the first point $p$ under the map in question, whose 
definition is now given by $H(p+v):=q+h(v)$. Two such maps, $H$ and $H'$,
characterised by $(p,q,h)$ and $(p',q',h')$, respectively, are 
easily seen to be the same iff $h=h'$ and $q'-q=h(p'-p)$. This defines
an equivalence relation on the set $\M\times\M\times\GL(\vecM)$,
the equivalence classes of which are 
\begin{equation}
\label{eq:AffineGroupEquivClasses}
[p,q,h]=\bigcup_{v\in\vecM}\Bigl(p+v,q+h(v),h\Bigr)\,.
\end{equation}
Hence we may identify $\Aff(\M)$ with this quotient space and 
write $H=[p,q,h]$ for any $H\in\Aff(\M)$. The composition of 
two maps $H=[p,q,h]$ and $H'=[p',q',h']$ can then be calculated 
\begin{equation}
\label{eq:AffineCompositions-1}
\begin{split}
H'\circ H(p+v)
&=H'\bigl(q+h(v)\bigr)=H'\bigl(p'+(q-p')+h(v)\bigr)\\
&=q'+h'(q-p')+h'\circ h(v)\,.
\end{split}
\end{equation}
In other words
\begin{equation}
\label{eq:AffineCompositions-2}
[p',q',h']\circ [p,q,h]=[p,q'+h'(q-p'),h'\circ h]\,.
\end{equation}
The first thing to note is that the equivalence class on the 
right-hand side is unchanged if we replace $(p,q,h)$ with 
$\bigl(p+v,q+h(v),h\bigr)$ or $(p',q',h')$ with $\bigl(p'+v',q+h'(v'),h'\bigr)$,
which means that this prescription written down in terms of 
representatives defines indeed a multiplication of equivalence 
classes. Note that the neutral element is $[p,p,\id_\vecM]$ 
and the inverse of $[p,q,h]$ is
\begin{equation}
\label{eq:AffInverse}
[p,q,h]^{-1}=[p,p-h^{-1}(q-p),h^{-1}]\,.
\end{equation}
Furthermore, it is easy to check that \eqref{eq:AffineCompositions-2}
is associative and hence defines a group multiplication. 

An obvious subgroup in $\Aff(\M)$ is given by the following 
subset
\begin{equation}
\label{eq:TranslationSubgroup}
\Trans(\M):=\bigl\{[p,q,h]\in\Aff(\M) \mid h=\id_\vecM\bigr\}\,.
\end{equation}
This subgroup is abelian,
\begin{equation}
\label{eq:TranslationsAreAbelian}
\begin{split}
[p',q',\id_\vecM]\circ [p,q,\id_\vecM]
&=[p,q'+(q-p'),\id_\vecM]\\
&=[p'+(p-p'),q'+(q-p'),\id_\vecM]\\
&=[p',q'+(q-p')+(p'-p),\id_\vecM]\\
&=[p',q'+(q-p),\id_\vecM]\\
&=[p',q+(q'-p),\id_\vecM]\\
&=[p,q,\id_\vecM]\circ [p',q',\id_\vecM]\,,
\end{split}
\end{equation}
(using \eqref{eq:MinusProperty-1} and \eqref{eq:MinusProperty-2}
at the fourth and fifth equality) and normal,
\begin{equation}
\label{eq:TranslationsAreNormal}
\begin{split}
&[p,q,h]\circ [p',q',\id_\vecM]\circ[p,q,h]^{-1}\\
&=[p,q,h]\circ [p',q',\id_\vecM]\circ[p,p-h^{-1}(q-p),h^{-1}]\\
&=[p,q,h]\circ [p,q'+(p-p')-h^{-1}(q-p),h^{-1}]\\
&=[p,q+h(q'-p')+(p-q),\id_\vecM]\\
&=[p,p+h(q'-p'),\id_\vecM]\,.
\end{split}
\end{equation}
It is called the \emph{subgroup of translations}. It is the 
kernel of the projection homomorphism 
\begin{equation}
\label{eq:ProjectitionHomomorphism}
\pi:\Aff(\M)\rightarrow\GL(\vecM)\,,\quad
[p,q,h]\mapsto\pi\bigl([p,q,h]\bigr):=h\,.
\end{equation}
If we denote the embedding (injective homomorphism) of 
$\Trans(\M)$ into $\Aff(\M)$ by $i$, we have the short 
sequence of groups and maps  
\begin{equation}
\label{eq:ShortExactSequence}
\bfig
\morphism(0,0)/->/<500,0>[\{1\}`\Trans(\M);{}]%
\morphism(500,0)|a|/ >->/<650,0>[\Trans(\M)`\Aff(\M);i]%
\morphism(1150,0)|a|/ ->>/<600,0>[\Aff(\M)`\GL(\vecM);\pi]%
\morphism(1750,0)/->/<450,0>[\GL(\vecM)`\{1\};{}]%
\efig
\end{equation}
Here $\{1\}$ stands for the trivial group with unique group 
homomorphims from and to any other group. The tailed and 
double-headed arrows indicate injective and surjective 
homomorphisms respectively. This may be briefly summarised 
by saying that the short sequence is \emph{exact}, where 
exactness means that at each group the image of the arriving 
map is the kernel of the departing one. 

Moreover, our sequence \eqref{eq:ShortExactSequence} is not 
only exact but it also \emph{splits}. By this is meant that 
there are also group embeddings (injective homomorphisms) 
$j:\bfig\morphism/ >->/<550,0>[\GL(\vecM)`\Aff(\M);{}]\efig$
so that $\pi\circ j=\id_{\GL(\vecM)}$. To see this, choose a point 
$o\in\M$ and define (indicating the dependence of $j$ on $o$ 
by a subscript) 
\begin{equation}
\label{eq:GL-Embedding}
j_o:
\bfig\morphism/ >->/<550,0>[\GL(\vecM)`\Aff(\M);{}]\efig
\,,\quad
h\mapsto j_o(h):=[o,o,h]\,.
\end{equation} 
Since $[o,o,h']\circ [o,o,h]=[o,o+h'(o-o),h'h]=[o,o,h'h]$ 
one has indeed $i_o(h')i_o(h)=i_o(h'h)$ and 
$i_o(\id_{\GL(\vecM)})=\id_{\Aff(\M)}$, that is, $i_o$ is a 
group homomorphism. But note that we needed to select a 
point $o\in\M$ to define the embedding. Two embeddings 
corresponding to different choices $o$ and $o'$ are 
related by conjugation with the translation from $o$ to 
$o'$. Indeed, using that according to \eqref{eq:AffInverse} 
we have $[o,o',\id_\vecM]^{-1}=[o,o-(o'-o),\id_\vecM]=[o',o,\id_\vecM]$,
we have for all $h\in\GL(\vecM)$
\begin{equation}
\label{eq:GL-Embedding-Covariance}
\begin{split}
[o,o',\id_\vecM]\circ i_o(h)\circ [o,o',\id_\vecM]^{-1}
&=[o,o',\id_\vecM]\circ [o,o,h]\circ [o',o,\id_\vecM]\\
&=[o,o',\id_\vecM]\circ [o',o,h]\\
&=[o',o',h]\\
&=i_{o'}(h)\,.
\end{split}
\end{equation} 
The relation between the three groups $\Trans(\M)$, $\Aff(\M)$, 
and $\GL(\vecM)$ can then be compactly expressed by completing 
the short exact sequence \eqref{eq:ShortExactSequence} by a  
splitting homomorphism $j_o$: 
\begin{equation}
\label{eq:ShortExactSequenceSplit}
\bfig
\morphism(0,0)/ ->/<500,0>[\{1\}`\Trans(\M);{}]%
\morphism(500,0)/ >->/<650,0>[\Trans(\M)`\Aff(\M);i]%
\morphism(1150,0)|a|/{@{->>}@/^5pt/}/<600,0>[\Aff(\M)`\GL(\vecM);\pi]%
\morphism(1150,0)|b|/{@{<-< }@/^-5pt/}/<600,0>[\Aff(\M)`\GL(\vecM);j_o]%
\morphism(1750,0)/ ->/<450,0>[\GL(\vecM)`\{1\};{}]%
\efig
\end{equation}
This characterisation in terms of a split exact-sequence is 
the most natural in view of the homogeneity of $M$. The usual 
characterisation by means of a semi-direct product $V\rtimes\GL(\vecM)$
is unnatural insofar as the $\GL(\vecM)$ subgroup in $\Aff(M)$ depends 
on the choice of a point $o\in M$, violating homogeneity. 
What one may say is that $\Aff(M)$ is isomorphic to $V\rtimes\GL(\vecM)$,
but the isomorphism depends on the selection of a point. Only 
\emph{after} the point is selected can we locate a linear
subgroup in $\Aff(M)$ isomorphic to $\GL(\vecM)$, namely the image 
of $\GL(\vecM)$ under the embedding $j_o$~\eqref{eq:GL-Embedding}.
Once one agrees to select a point $o\in M$, we may write the 
general element of $\Aff(M)$ in the form $[o,q,h]$. Group 
multiplication according to \eqref{eq:AffineCompositions-2} 
then becomes
\begin{equation}
\label{eq:AffineCompositions-alt}
[o,q',h']\circ [o,q,h]
=[o,q'+h'(q-o),h'\circ h]
=[o,o+(q'-o)+h'(q-o),h'\circ h]\,.
\end{equation}
Having selected $o$ we may identify $M$ with $V$ via 
$p\mapsto p-o$ (sometimes called the ``vectorialisation''
of $M$ at $o$ \cite{Berger:Geometry1}) and the group 
$\Aff(M)$ with the set $\vecM\times\GL(\vecM)$. A general 
group element may then be written $[o,o+v,h]\mapsto(v,h)$ 
and \eqref{eq:AffineCompositions-alt} becomes 
\begin{equation}
\label{eq:AffineCompositions-alt-2}
(v',h')\circ (v,h)=\bigl(v'+h'(v),h'\circ h\bigr)\,,
\end{equation}
which is just the product structure of a semi-direct
product $\vecM\times\GL(\vecM)$ with respect to the homomorphism 
$\GL(\vecM)\rightarrow\Aut(\vecM)$ that is given by the defining 
representation of $\GL(\vecM)$.

\begin{remark}
\label{rem:SemiDirektproduct}
The proper statement regarding the structure of the affine 
group $\Aff(M)$ is that it is a \emph{downward splitting extension}
of $\GL(\vecM)$ by $\Trans(M)$, as summarised by 
\eqref{eq:ShortExactSequenceSplit}. To be a downward extension%
\footnote{Here we recall that the usual terminology regarding 
extensions of groups is not quite uniform and hence ambiguous. 
Suppose three groups $H,E$ and $G$ are related by an exact 
sequence $1\rightarrow H\rightarrow E\rightarrow G\rightarrow 1$,
i.e. that $H$ is a normal (or ``invariant'') subgroup of $E$ with quotient 
$E/H$ isomorphic to $G$. Then this state of affairs is 
usually simply expressed by \emph{either} saying that $E$ is 
``an extension'' of $G$ by $H$, \emph{or} of $H$ by $G$. 
This ambiguity arises because views differ as to whether 
one likes to regard the \emph{extending} or the \emph{extended}
group to be that one which becomes normal in the extension.
To avoid such ambiguities the following refined terminology 
has been proposed in \cite{Atlas:FiniteGroups}: $E$ is 
called an \emph{upward extension} of $H$ by $G$, or a 
\emph{downward extension} of $G$ by $H$.} 
means that $\Trans(M)$ is a normal (or ``invariant'') subgroup of $\Aff(M)$ so 
that the quotient $\Aff(M)/\Trans(M)$ is isomorphic to $\GL(\vecM)$. 
To be ``splitting'' means that $\GL(\vecM)$ may be identified 
with a subgroup in $\Aff(\M)$ whose intersection with $\Trans(\M)$
is merley the group identity. In our case there exist many 
such splitting embeddings of $\GL(\vecM)$ into $\Aff(\M)$, so 
that there is no unique way to regard $\GL(\vecM)$ as subgroup 
of $\Aff(M)$. The ambiguity is faithfully labelled by the points 
in $\M$ (the point that is fixed under the action of the 
embedded copy of $\GL(\vecM)$ in $\Aff(M)$ on $\M$). Given such 
a splitting, $\Aff(M)$ becomes isomorphic to the corresponding
semi-direct product $\vecM\rtimes\GL(\vecM)$. But this isomorphism 
depends on the choice of a point in $\M$. If one says that 
$\Aff(M)$ is isomorphic to  $\vecM\rtimes\GL(\vecM)$ one should add
that this isomorphisms is not ``natural'', since by the very 
homogeneity of $\M$ there is clarly no preferred choice of a 
point in $\M$. 
\end{remark}

\subsection{Poincar\'e group}
\label{eq:Poinvare-Group}
\begin{definition}
\label{def:PoincareGroup}
Given Definition\,\ref{def:MinkowskiSpace} of Minkowski space, 
we define the \emph{Poincar\'e group}, $\Poin(\M)$, to be its 
group of automorphisms. This means that is must consists of 
affine transformations including all elements in $\Trans(\M)$,
such that $\Aff(\M)/\Trans(\M)=\Lor(\vecM)\subset GL(\vecM)$, 
where 
\begin{equation}
\label{def:LorentzGroup}
\Lor(\vecM):=\Bigl\{h\in\GL(\vecM)\mid\eta\bigl(h(v),h(w)\bigr)=\eta\bigl(v,w\bigr)\,,\forall v,w\in V\Bigr\}\,.
\end{equation}
Hence we have 
\begin{equation}
\label{eq:DefPoincaregroup}
\Poin(\M):=
\bigl\{H:\M\rightarrow\M \mid H(p+v)=H(p)+h(v),\, h\in\Lor(\vecM)\,,\forall v\in\vecM\bigr\}\,.
\end{equation}
\qed
\end{definition}

Totally analogous to \eqref{eq:ShortExactSequenceSplit},
this leads to the splitting exact sequence  
\begin{equation}
\label{eq:ShortExactSequenceSplitPoin}
\bfig
\morphism(0,0)/ ->/<500,0>[\{1\}`\Trans(\M);{}]%
\morphism(500,0)/ >->/<650,0>[\Trans(\M)`\Poin(\M);i]%
\morphism(1150,0)|a|/{@{->>}@/^5pt/}/<600,0>[\Poin(\M)`\Lor(\vecM);\pi]%
\morphism(1150,0)|b|/{@{<-< }@/^-5pt/}/<600,0>[\Poin(\M)`\Lor(\vecM);j_o]%
\morphism(1750,0)/ ->/<450,0>[\Lor(\vecM)`\{1\};{}]%
\efig
\end{equation}
and to the $o\in\M$ dependent(!) isomorphism 
\begin{equation}
\label{eq:PoincareSemiDirect}
\Poin(\M)\cong V\rtimes\Lor(\vecM)\,.
\end{equation}
If we complete $o$ to a full affine frame $F=(o,f)$, where 
$f\in\Lin(\reals^n,V)$, and if in addition we require $f$ to map the 
standard basis of $\reals^n$ to the orthonormal basis of $V$ 
with respect to $\eta$, i.e., $\eta_{ab}:=\eta(e_a,e_b)=\pm\delta_{ab}$ 
with one plus and $n-1$ minus signs), we may identify $\M$ with 
$\reals^n$, and then have 
\begin{equation}
\label{eq:PoincareSemiDirect-alt}
\Poin(\reals^n)=\reals^n\rtimes\Lor(\reals^n)\,.
\end{equation}
where 
\begin{equation}
\label{def:LorentzGroupMatrices}
\Lor(\reals^n):=\bigl\{L\in\GL(R^n) \mid \eta_{ab}L^a_cL^b_d=\eta_{cd}\bigr\}\,.
\end{equation}

Now, $\Poin(\M)$ is a Lie group. The structure of a differentiable 
manifold with respect to which all group operations become smooth 
are again obtained by its isomorphism (non-naturalness is irrelevant
here) with the matrix group just described. Note that the 
semi-direct product \eqref{eq:PoincareSemiDirect-alt} can itself 
be embedded (i.e. mapped by an injective homomorphism) into 
the group $\GL(\reals^{n+1})$, via
\begin{equation}
\label{eq:PoinIntoGenLin}
\reals^n\rtimes\Lor(\reals^n)\ni(v,L)\mapsto
\begin{pmatrix}
1&0\\
v&L\\
\end{pmatrix}\in\GL(\reals^{n+1})
\end{equation}
which endows it with the differentiable structure inherited from 
$\GL(\reals^{n+1})$. All this is using the preferred affine (or inertial) 
coordinates of $\M$; compare Definition\,\ref{def:AffineCoordinates}. 
Note also that the group multiplication in $\Aff(\M)$ has been 
explained simply by composition of maps ($\Aff(\M)$ was defined to 
consists of special bijections of $\M$). This defines a left 
action of $\Aff(\M)$ on $\M$ and hence, by simple restriction, 
a left action of $\Poin(\M)$ on $\M$.  This, in turn, defines an 
anti-homomorphism between the Lie -algebra of $\Poin(\M)$ and the 
Lie algebra of vector fields on $\M$ (considered as differentiable 
manifold), where the Lie-algebra  structure of the latter is 
defined by the commutator of vector fields. The reason why we have 
an anti- rather than a proper homomorphism of Lie algebras is 
explained in detail in Appendix\,\ref{sec:Appendix-GroupsActions}, 
in which we also review in some detail the notion of Lie-group actions 
on manifolds. 

We recall that the Lie algebra, $\LieLor(\vecM)$, of $\Lor(\vecM)$ 
is the linear space of endomorphisms $A\in\End(\vecM)$ which are 
antisymmetric with respect to the Minkowski inner product $\eta$,
i.e., satisfy $\eta(Av,w)=-\eta(v,Aw)$ for all $v,w\in\vecM$. 
Using the $\eta$-induced isomorphism $\etadown:\vecM\rightarrow
\vecM^*$ and its inverse $\etaup$ (compare 
Remark\,\ref{rem:MetricAndNotation}), we can then write down 
the projection 
operators $P_S,P_A:\End(\vecM)\rightarrow\End(\vecM)$, which 
project onto the $\eta$-symmetric and $\eta$-antisymmetric 
endomorphisms: 
\begin{subequations}
\label{eq:ProjectorsInEnd}
\begin{alignat}{2}
\label{eq:ProjectorsInEnd-a}
&P_S(M)&&\,:=\,\tfrac{1}{2}\bigl(M+\etaup\circ M^\top\circ\etadown\bigr)\,,\\
\label{eq:ProjectorsInEnd-b}
&P_A(M)&&\,:=\,\tfrac{1}{2}\bigl(M-\etaup\circ M^\top\circ\etadown\bigr)\,.
\end{alignat}
\end{subequations}
Hence $\LieLor(\vecM)$ can be either characterised as the kernel of 
$P_S$ or the image of $P_A$ in $\End(\vecM)$. Using the first option 
we may write 
\begin{equation}
\label{def:LorentzLieAlgebra}
\LieLor(\vecM)=\kernel(P_S)=
\bigl\{A\in\End(\vecM)\mid A=-\etaup\circ A^\top\circ\etadown\bigr\}\,.
\end{equation}

Using the point-dependent isomorphism~\eqref{eq:PoincareSemiDirect}, 
the Lie algebra of $\Poin(\M)$, denoted by $\LiePoin(\M)$, is the 
semi-direct product of the Lie algebras $\vecM$ and $\LieLor(\vecM)$.
Note that $\vecM$, considered as abelian group, has a Lie algebra which
is isomorphic (as vector space) to $\vecM$ with trivial Lie product 
(i.e. all Lie products are zero). Then we get the, likewise point-dependent, 
isomorphism 
\begin{equation}
\label{eq:LiePoincareSemiDirect}
\begin{split}
&\LiePoin(\M)\cong\vecM\rtimes\LieLor(\vecM)\\
&=\Bigl\{
(v,M)\in\vecM\times\LieLor(\vecM) \mid
\bigl[(v,M),(w,N)\bigr]=\bigl(Mw-Nv\,,\,[M,N]\bigr)
\Bigr\}\,.
\end{split}
\end{equation}
Here $Mw$ is the action of $M\in\End(\vecM)$ on $w\in\vecM$
and $[M,N]$ is the commutator, which turns $\End(\vecM)$,
considered as associative algebra, into a Lie algebra. 
An easy way to see that \eqref{eq:LiePoincareSemiDirect} 
does indeed give the right Lie product is to use the 
embedding \eqref{eq:PoinIntoGenLin}, which induces an 
embedding   
\begin{equation}
\label{eq:LiePoinIntoGenEnd}
\reals^n\rtimes\LieLor(\reals^n)\ni(v,M)\mapsto
\begin{pmatrix}
0&0\\
v&M\\
\end{pmatrix}\in\End(\reals^{n+1})\,.
\end{equation}
The Lie product of the images of $(v,M)$ and $(w,N)$
is then just their commutator, which is immediately seen 
to be the image of $\bigl(Mw-Nv,[M,N]\bigr)$.

Now, as already mentioned above, the left action 
\begin{equation}
\label{eq:LeftAction}
\Phi: \Poin(\M)\times\M\rightarrow\M\,,\quad
(g,m)\mapsto\Phi(g,m)\equiv\Phi_g(m)
\end{equation}
of $\Poin(\M)$ on $\M$ induces a linear map from $\LiePoin(\M)$ 
to the linear space of vector fields on $\M$, denoted by $\Vect(M)$
(smooth sections in $TM$).  
This map is just the differential of $\Phi$ with respect to the 
first (group valued) argument 
evaluated at the group identity. This is explained in all detail 
in Appendix\,\ref{sec:Appendix-GroupsActions}; compare~\eqref{eq:GroupAction_FundVecField}.
Since $\Vect(\M)$  is itself a Lie algebra, where the Lie 
product is defined to be the commutator of vector fields. 
With respect to these two Lie structures, the linear map 
$\LiePoin(\M)\ni X\mapsto V^X\in\Vect(M)$ is a Lie 
anti-homomorphism. Again we refer to the  
Appendix\,\ref{sec:Appendix-GroupsActions} for details;  
compare~\eqref{eq:GroupAction_GenCommRightLeftAction-b}. 
Hence we have     
\begin{equation}
\label{eq:LieAntiHomom}
\bigl[V^X,V^Y\bigr]=-V^{[X,Y]}\,, 
\end{equation}
where the ``anti'' is reflected by the minus-sign on the 
right-hand side. 

Moreover, as the left action of $\Poin(\M)$ on $\M$ lifts 
by push-forward (differential of $\Phi$ with respect to 
second ($\M$-valued) argument) to a left action on $T\M$ 
and hence  $\Vect(\M)$, we can ask for the result of acting 
with $g\in\Poin(\M)$ on the special vector field $V^X$. 
The result is (see 
equation~\eqref{eq:GroupActions_AdEquivLeftRightaction-a} 
of Appendix\,\ref{sec:Appendix-GroupsActions})
\begin{equation}
\label{eq:PushForwardFundVecField}
\Phi_{g*}V^X=V^{\Ad_g(X)}\circ\Phi_g\,.
\end{equation}
where $\Ad$ denotes the adjoint representation of $\Poin(\M)$
on $\LiePoin(\M)$. 

Let us at this point say a few words about the adjoint and co-adjoint 
representation; the latter will become important in what is to follow. 
An easy way to calculate the adjoint representation is again to 
identify $\Poin(\M)$ and $\LiePoin(\M)$ according to 
\eqref{eq:PoincareSemiDirect-alt} and \eqref{eq:LiePoincareSemiDirect}, 
respectively, and perform the easy conjugation-calculation using the 
embeddings \eqref{eq:PoinIntoGenLin} and \eqref{eq:LiePoinIntoGenEnd}. 
The result is 
\begin{equation}
\label{eq:AdjointRepPoin}
\Ad_{(a,L)}(v,M)=\bigl(Lv-LML^{-1}v\,,\,LML^{-1}\bigr)\,.
\end{equation}
The co-adjoint representation is the usual representation 
induced by $\Ad$ on the dual space, that is, the inverse 
transposed. As a vector space, $\LiePoin(\M)$ is isomorphic 
to a linear subspace of $\vecM\oplus\End(\vecM)$, namely the image of 
$\id_\vecM\oplus P_A$. Note that $\vecM\oplus\End(\vecM)$ may 
be identified with $\vecM\oplus(\vecM\otimes \vecM^*)$. The dual of the vector 
space $\LiePoin(\M)$ is then isomorphic to a subspace of the dual to 
$\vecM\oplus(\vecM\otimes\vecM^*)$, i.e., a subspace of 
$\vecM^*\oplus(\vecM^*\otimes\vecM)$. This subspace is the image of 
$\id_{V^*}\oplus P_A^\top$. It is called the dual of the 
Lie algebra $\LiePoin(M)$, denoted by $\LiePoin^*(M)$. It is 
merely considered as a vector space, not a Lie algebra. The natural 
paring between $(p,J)\in\LiePoin^*(\M)$ and $(v,M)\in\LiePoin(\M)$ is 
\begin{equation}
\label{eq:NaturalPairingLiePoin}
[(p,J)](v,M)=p(v)+\tfrac{1}{2}\trace(J^\top\circ M)\,.  
\end{equation}
The factor $1/2$ in the second term is introduced because 
$M$ obeys the condition $P_S(M)=0$ and each independent 
component of $M$ contributes twice to the trace. We have,
by definition of the transposed map, 
$\trace(J^\top\circ P_AM)=:\trace((P^\top_AJ)^\top\circ M)$ and 
likewise for $P^\top_S$, which immediately leads to the expressions 
\begin{subequations}
\label{eq:ProjectorsInEndStar}
\begin{alignat}{2}
\label{eq:ProjectorsInEndStar-a}
&P^\top_S(J)&&\,:=\,\tfrac{1}{2}\bigl(J+\etadown
\circ J^\top\circ\etaup\bigr)\,,\\
\label{eq:ProjectorsInEndStar-b}
&P^\top_A(J)&&\,:=\,\tfrac{1}{2}\bigl(J-\etadown\circ J^\top\circ\etaup\bigr)\,.
\end{alignat}
\end{subequations}
Hence we may characterise $\LieLor^*(\vecM)$ by:
\begin{equation}
\label{eq:ProjectionLieLorDual}
\LieLor^*(\vecM)=\bigl\{J\in\End(\vecM^*) \mid J=-\etadown\circ J^\top\circ\etaup\bigr\}\,,
\end{equation}
and furthermore (as vector spaces) 
\begin{equation}
\label{eq:DualLiePoincare}
\LiePoin^*(\M)\cong \vecM^*\times\LieLor^*(\vecM)\,.
\end{equation}
As already said, the co-adjoint representation, $\Ad^*$ 
of $\Poin(\M)$ on $\LiePoin^*(\M)$ is defined to be the 
inverse-transposed:
\begin{equation}
\label{eq:CoAdjointRep}
\begin{split}
\Ad^*:\Poin(\M)\times\LiePoin^*(\M)
&\rightarrow\LiePoin^*(\M)\\
\bigl((a,L)\,,\,(p,J)\bigr)
&\mapsto \Ad^*_{(a,L)}(p,J):=(p,J)\circ\Ad^{-1}_{(a,L)}\,.
\end{split}
\end{equation}
Note that the inverse is necessary to get a left action, i.e.,  
$\Ad^*_{(a,L)}\circ\Ad^*_{(a',L')}=\Ad^*_{(a,L)(a',L')}$. 
Using 
\begin{equation}
\label{eq:InverseAdjoint-1}
\Ad^{-1}_{(a,L)}(v,M)=\bigl(L^{-1}v+L^{-1}Ma\,,\,L^{-1}ML\bigr)\,,
\end{equation}
a straightforward calculation gives, writing $\tilde L:=(L^\top)^{-1}$
and using the identity $p(w)=\trace(w\otimes p)$, valid for any $w\in\vecM$
and $p\in\vecM^*$,  
\begin{equation}
\label{eq:InverseAdjoint-2}
[\Ad^*_{(a,L)}(p,J)](v,M)=\tilde Lp(v)
+\tfrac{1}{2}\trace\Bigl(\bigl[2\,\tilde Lp\otimes a+\tilde LJ\tilde L^{-1}\bigr]^\top M\Bigr)\,.
\end{equation}
This implies 

\begin{equation}
\label{eq:CoAdjointRepPoin}
\begin{split}
\Ad^*_{(a,L)}(p,J)
&=\bigl(\tilde Lp\,,\,\tilde LJ\tilde L^{-1}+P_A^\top(2\,\tilde Lp\otimes a\bigr)\\
&=\bigl(\tilde Lp\,,\,\tilde LJ\tilde L^{-1}+\tilde Lp\otimes a-a^\flat\otimes (\tilde Lp)^\sharp\bigr)\,.
\end{split}
\end{equation}

For what follows it is important to compare the adjoint 
representation \eqref{eq:AdjointRepPoin} of $\Poin(\M)$ on 
$\LiePoin(\M)$ with the co-adjoint representation 
\eqref{eq:CoAdjointRepPoin}  of the same group on 
$\LiePoin^*(\M)$. This is not quite straightforward since 
the representation spaces are different and hence it is not 
entirely obvious how to best appreciate their difference. 
However, it is true that, as vector spaces, $\LiePoin(\M)$ 
and $\LiePoin^*(\M)$ are isomorphic, though not naturally so. 
We need an extra structure to select a specific isomorphism, 
which in our case is already given to us by the inner 
product $\eta$, which was already seen to give an isomorphism 
$\etadown:\vecM\rightarrow\vecM^*$; compare Remark\,\ref{rem:MetricAndNotation}. This structure can clearly also be 
used to define an isomorphisms  $\LiePoin(\M)\rightarrow\LiePoin^*(\M)$. 
However, it is more convenient to define isomorphisms between each of 
these vector spaces and $\vecM\oplus\bigwedge^2V$, where
$\bigwedge^2V:=V\wedge V$ is the antisymmetric tensor product: 
\begin{equation}
\label{eq:IsomLiePoinAndDual}
\LiePoin(\M)
\cong\vecM\oplus P_A(\vecM\otimes\vecM^*)
\cong\vecM\oplus(\vecM\wedge\vecM)
\cong\vecM^*\oplus P_A^\top(\vecM^*\otimes\vecM)
\cong\LiePoin^*(\M)\,.
\end{equation}
Indeed, note that under this isomorphism $\LieLor(\vecM)$ 
gets mapped isomorphically onto the antisymmetric 
subspace $\bigwedge^2V\subset V\otimes V$. The 
corresponding representations on $V\oplus\bigwedge^2V$,
which are equivalent to $\Ad$ and $\Ad^*$ under these 
isomorphisms, are respectively given by
\begin{subequations}
\label{eq:AdjointCoAdjointReps}
\begin{alignat}{2}
\label{eq:AdjointCoAdjointReps-a}
&Ad_{(a,L)}(v,M)&&\,=\,\bigl(Lv-(L\otimes LM)\cdot a\,,\,L\otimes LM\bigr)\,,\\
\label{eq:AdjointCoAdjointReps-b}
&Ad^*_{(a,L)}(p,J)&&\,=\,\bigl(Lp\,,\,L\otimes LJ-a\wedge Lp\bigr)\,,
\end{alignat}
\end{subequations}
where the dot now abbreviates the inner product $\eta$ 
in $V$, as explained in Remark\,\ref{rem:MetricAndNotation}. 
So for $a,b,c\in\vecM $ we write 
\begin{equation}
\label{eq:DefDotNotation} 
(a\wedge b)\cdot c
=(a\otimes b-b\otimes a)\cdot c
=a\,\eta(b,c)-b\,\eta(a,c)
=:a (b\cdot c)-b (a\cdot c)\,.
\end{equation}
These are now two inequivalent representations of the same 
group on the same vector space. It is the second, co-adjoint 
representation that will be physically relevant. It differs 
from the adjoint representation on how it implements the 
normal subgroup of translations. Let us, for clarity, just 
display the two representations if restricted to the subgroup 
$\Trans(\M)$:
\begin{subequations}
\label{eq:AdjointCoAdjointRepsTrans}
\begin{alignat}{2}
\label{eq:AdjointCoAdjointRepsTrans-a}
&Ad_{(a,\id)}(v,M)&&\,=\,\bigl(v-M\cdot a\,,\,M\bigr)\,,\\
\label{eq:AdjointCoAdjointRepsTrans-b}
&Ad^*_{(a,\id)}(p,J)&&\,=\,\bigl(p\,,\,J-a\wedge p\bigr)\,.
\end{alignat}
\end{subequations}
The obvious difference is that under the adjoint representation 
translations act non-trivially only on the first summand in $\vecM\oplus(\vecM\wedge\vecM)$ under the adjoint-, and 
non-trivially only on the second summand under the co-adjoint 
representation. As we will see below, the latter corresponds 
to the familiar origin-dependence of angular momentum and 
origin-independence of linear momentum.   

Finally, using the identification 
$\Lie\Poin(\M)\cong\vecM\rtimes(\vecM\wedge\vecM)$, 
let us explicitly write down the Lie algebra 
\eqref{eq:LiePoincareSemiDirect} in terms of a basis.
Let $\{e_a\mid 1\leq a\leq n\}$ be a  basis of $\vecM$, such 
that $\eta(e_a,e_b)=e_a\cdot e_b=\eta_{ab}$ then 
$\{m_{ab}\mid 1\leq a<b\leq n\}$ is a basis 
of $\vecM\wedge\vecM$, where 
$m_{ab}:=e_a\wedge e_b=(e_a\otimes e_b-e_b\otimes e_a)$. 
Then the Lie products in \eqref{eq:LiePoincareSemiDirect}
become
\begin{subequations}
\label{eq:LiePoinIn Basis}
\begin{alignat}{2}
\label{eq:LiePoinIn Basis-a}
&[e_a,e_b]&&\,=\,0\,,\\
\label{eq:LiePoinIn Basis-b}
&[e_a,m_{bc}]&&\,=\,\eta_{ab}\,e_c-\eta_{ac}\,e_b\,,\\
\label{eq:LiePoinIn Basis-c}
&[m_{ab},m_{cd}]&&\,=\,
\eta_{ad}\,m_{bc}+
\eta_{bc}\,m_{ad}-
\eta_{ac}\,m_{bd}-
\eta_{bd}\,m_{ac}\,.
\end{alignat} 
\end{subequations}

 \section{The momentum map and the natural habitat 
of globally conserved Poincar\'e charges}
\label{sec:MomentumMap}
We now regard Minkowski space as a Semi-Riemannian manifold $(M,g)$
with Lorentzian metric $g$, which in affine/inertial coordinates
(compare Definition\,\ref{def:AffineCoordinates}) is of the form 
(in four spacetime-dimensions)
\begin{equation}
\label{eq:MinkowskimetrikOnMannifold}
g=\eta_{ab}\ dx^a\otimes dx^b\,,\quad
\{\eta_{ab}\}=\mathrm{diag}(1,-1,-1,-1)\,.
\end{equation}
As discussed above, and in more detail in the 
Appendix\,\ref{sec:Appendix-GroupsActions}, for each 
$X\in\LiePoin(\M)$ we have a vector field $V^X\in\Vect(\M)$ 
that represents the ``infinitesimal'' left group-action of 
$\Poin(\M)$ on $\M$ through an anti Lie-homomorphism 
$\LiePoin(\M)\rightarrow\Vect(\M)$, $X\mapsto V^X$, satisfying 
\eqref{eq:LieAntiHomom}. Since $\Poin(\M)$ acts on $\M$ by
isometries, the Lie derivative of $g$ with respect to each $V^X$
is zero:
\begin{equation}
\label{eq:FundVectAreKilling}
L_{V^X}g=0\,.
\end{equation} 
In other words, each $V^X\in\Vect(\M)$ is a Killing vector-field.

Now, suppose we have an energy-momentum tensor
\begin{equation}
\label{eq:EnergyMomentumtensor}
\EMT=\EMT_{ab}\ dx^a\otimes dx^b
\end{equation}
which is divergence free with respect to the Levi-Civita 
covariant derivative determined by $g$. In components with 
respect to arbitrary coordinate systems this reads 
\begin{equation}
\label{eq:ZeroDivEMT}
\nabla_aT^{ab}=\partial_aT^{ab}+\Gamma^a_{ac}T^{cb}+\Gamma^b_{ac}T^{ac}=0\,.
\end{equation}
If the coordinates are affine/inertial, the $\Gamma$-coefficients 
are all zero. 

Another way to look at $\EMT$ is to regard it as a co-vector valued 
3-form. This is achieved by Hodge dualising the second tensor
factor in \eqref{eq:EnergyMomentumtensor}: 
\begin{equation}
\label{eq:EMTvv3Form}
\EMTForm = T_{ab}\ dx^a\otimes (\star dx^b)\,,
\end{equation}  
where $\star$ is the Hodge duality map the definition of which, 
together with our conventions, are summarised in Appendix\,\ref{sec:Appendix-ExtProdHodge}. Now comes the important 
point in the whole construction: using the vector fields $V^X$,
we can, for each $X\in\LiePoin(\M)$ turn \eqref{eq:EMTvv3Form}
into a 3-form that linearly depends on $X$ via 
\begin{equation}
\label{eq:EMT3Form}
\EMTForm_X:
=\star\,i_{V^X}\EMT 
=(V^X)^aT_{ab}\, (\star dx^b)
=(V^X)^a\,g_{ab}\,T^{bc}\,\tfrac{1}{3!}\varepsilon_{cdef}dx^d\wedge dx^e\wedge dx^f\,.
\end{equation}  
here $i_{V}$ denotes the map of inserting $V$ into the first 
co-vector factor of the tensor it is applied to and 
$\varepsilon_{abcd}$ are the components of the measure 4-form 
induced by $g$. The zero-divergence condition 
\eqref{eq:ZeroDivEMT} implies, in view of 
\eqref{eq:FundVectAreKilling}, that each $\EMTForm_X$ is closed:
\begin{equation}
\label{eq:EMT3FormClosed}
d\EMTForm_X=0\,.
\end{equation}  
This means that to each $X\in\LiePoin(\M)$ we can produce a number 
by integrating $\EMTForm_X$ over a 3-dimensional hypersurface:
\begin{equation}
\label{eq:MomentumMap}
\MM[F,S](X):=\int_S\EMTForm_X[F]=\int_S \star\circ i_{V^X}\circ\EMT [F]\,.
\end{equation}
Here we wrote the integrand as a composition of three maps.
The first ($\EMT$) maps the field configuration $F$ to a 
symmetric tensor, the second $i_{V^x}$ contracts this tensor
with the vector field $V^X$ and turns it into a one form, 
and the last $(\star)$ turns this one form into an $n-1$
form (a three-form in four dimensions). The last map to 
be applied in order to get a number is to integrate this 
form over a hypersurface $S$. This number will depend on 
three arguments: The fields $F$ on which $\EMT$ depends, 
the surface $S$ over which we integrate, and the Lie algebra 
element $X$ which we use to build $V^X$ to contract $\EMT$ with. 
The value $\MM$ takes on all these arguments is called the 
corresponding \emph{momentum}. 

Suppose now that the fields $F$ on which $\EMT$ depends 
carry a representation (not necessarily a linear one) of 
$\Poin(\M)$. That is, we assume there is a left action $D$ 
of $\Poin(\M)$ on the space (not necessarily a vector space)
of fields. We assume that the geometric object $\EMT$ is built 
entirely out of such fields, and that there is no dependence 
on any other geometric structure not included in our $F$. 
Then we have the covariance property\footnote{This covariance 
property, which is crucial for the right representation-theoretic 
properties of the global charges, is hardly ever stated explicitly. 
A notable exception, more in words than in formulae, is Fock's
book \cite{Fock:STG} \S\,31, where it is refered to as ``physical 
principle''.} 
\begin{equation}
\label{eq:CovarianceEMT}
\EMT[D_hF]=\Phi_{h*}\EMT[F]
\end{equation}  
where $\Phi$ is as in \eqref{eq:LeftAction} and 
$\Phi_{h*}$ denotes the push-forward of the diffeomorphism 
$\Phi_h:\M\rightarrow\M$. If $F$ denote standard
scalar, vector, and tensor fields, then \eqref{eq:CovarianceEMT}
merely says that the energy-momentum distribution of the 
pushed-forward fields is just the push-forward of the 
energy-momentum distribution of the original fields.

Now we are interested in how the momentum changes if we 
act on the fields $F$ by a Poincar\'e transformation, 
leaving the arguments $S,X$ untouched for the moment. 
We get:

\begin{equation}
\begin{split}
\label{eq:TranfOfMomentum}
\MM[D_h(F),S](X)\
&\stackrel{ }{=}\ \int_S \star\circ i_{V^X}\circ T\circ D_h\ [F]\\
&\stackrel{1}{=}\ \int_S \star \circ i_{V^X}\circ \Phi_{h*}\circ T\ [F]\\ 
&\stackrel{2}{=}\ \int_S \star\circ \Phi_{h*}\circ i_{\Phi^{-1}_{h*}V^X}\circ T\ [F]\\
&\stackrel{3}{=}\ \int_S \star\circ \Phi_{h*}\circ i_{V^{\Ad_{h^{-1}}(X)}}\circ T\ [F]\\
&\stackrel{4}{=}\ \int_S \Phi^*_{h^{-1}}\Bigl(\star\circ  i_{V^{\Ad_{h^{-1}}(X)}}\circ T\ [F]\Bigr)\\
&\stackrel{5}{=}\ \int_{\Phi_{h^{-1}}(S)} \Bigl(\star\circ  i_{V^{\Ad_{h^{-1}}(X)}}\circ T\ [F]\Bigr)\\
&\stackrel{6}{=}\ \MM[F,\Phi_{h^{-1}}(S)](\Ad_h^{-1}(X))\\
&\stackrel{7}{=}\ \Ad^*_h(\MM)[F,\Phi_{h^{-1}}(S)](X)
\end{split}
\end{equation}
Here we broke up the derivation into seven steps, each one
showing what happens as we commute the action of $\Poin(\M)$ 
from right to left through the various maps connected by the 
$\circ$ symbols.  At the first step we use 
\eqref{eq:CovarianceEMT}, at the second step we just use the 
obvious commutation property of push-forwards with the 
vector-insertion map, at the third step we use property 
\eqref{eq:PushForwardFundVecField}, at the fourth step we 
use the covariance (intertwining property) of the Hodge map 
and the definition of the push-forward of a form as the 
pull-back by the inverse map, in the fifth step we use 
the elementary property of integrals, sometimes referred to as the 
``change-of-variables-formula'', in the sixth step 
we just use the definition \eqref{eq:MomentumMap}, and in the 
seventh and last step we use the definition \eqref{eq:CoAdjointRep} 
of the co-adjoint representation.

Now, if $S$ is a Cauchy surface and the support conditions discussed 
initially are satisfies, we are ensured that the integral converges 
and the momentum actually exists. Moreover, if $\EMT$ is divergence 
free, as we assume here, the momentum does not depend on the 
particular Cauchy surface chosen, as follows  follows from 
\eqref{eq:EMT3FormClosed} and Gauss' theorem. Hence we may delete 
$S$ as an argument of $\MM$. Since equation \eqref{eq:TranfOfMomentum}
is valid for all $X\in\LiePoin(\M)$, we may also delete the 
dependence on $X$, which is linear. We can then  and regard 
\eqref{eq:TranfOfMomentum} as an equation between elements in the 
dual of the Lie algebra depending merely on $F$ and expressing the 
fact that they transform under the co-adjoint representation.

\begin{theorem}
\label{def:MomentumMap}
A divergence-free energy-momentum tensor describing a body 
in the sense of Definition\,\ref{def:Body} and depending on 
fields which carry a (not necessarily linear) representation 
$D$ of $\Poin(\M)$ defines a map from the space of field 
configurations to $\LiePoin^*(\M)$, called \emph{momentum map},
given by 
\begin{equation}
\label{eq:DefMomentumMap}
\MM(X):= \int_S\EMTForm_X[F]\,,
\end{equation}
where $S$ is any Cauchy surface. 
The map is $\Ad^*$-\emph{equivariant} in the sense that 
\begin{equation}
\label{eq:MomentumMapAdEquiv}
\MM\circ D_h=\Ad_h^*\circ\MM\,,
\end{equation}
for all $h\in\Poin(\M)$.
\qed
\end{theorem}

Let us finally see how, and in what sense, the general 
formula \eqref{eq:DefMomentumMap} implies the naive 
expressions \eqref{eq:NaiveCharges}. For this we express 
$V^X$ in affine/inertial coordinates and choose for 
$X$ basis elements of $\LiePoin(\M)$ that are adapted 
to the decomposition of $\LiePoin(\M)$ as semi-direct 
product $\vecM\rtimes\LieLor(\vecM)$. But here comes the point 
stressed above: there is no \emph{natural} identification 
of $\vecM\rtimes\LieLor(\vecM)$ with $\LiePoin(\M)$. Any such 
identification is equivalent to the choice of a point $o\in\M$.
Only with respect to the choice of such a point does it make 
sense to speak of $\Lor(\vecM)$ as a subgroup of $\Poin(\M)$ 
and of $\LieLor(\vecM)$ as a Lie subalgebra of $\LiePoin(\M)$. 

Let us now choose a system $x^a$ of  affine/inertial coordinates
so that the vector fields $\partial/\partial x^a$ are orthonormal
(i.e. the Minkowski metric $g$ takes the standard form \eqref{eq:MinkowskimetrikOnMannifold}). The coordinate values 
of the preferred point $o$ is denoted by $z^a:=x^a(o)$. Then
$V^X$ for $X=(v,M)\in\vecM\oplus(\vecM\otimes\vecM)$ is 
\begin{equation}
\label{eq:FundVF_Coordinates}
V^{(v,M)}(z)=v^a\,\partial/\partial x^a
+\tfrac{1}{2}\,M^{ac}\eta_{cb}\bigl[
(x^a-z^a)\partial/\partial x^b-
(x^b-z^b)\partial/\partial x^a\bigr]\,.
\end{equation}
Note that $x^a:\M\rightarrow\reals$ are coordinate functions 
on the manifold whereas $z^a=x^a(o)$ are fixed numbers 
(constant functions on $\M$). The corresponding \emph{momentum}
is then      
\begin{equation}
\label{eq:MomentumCoordinates}
\MM\bigl[X=(v,M)\bigr]=\eta_{ab}v^aP^b
+\tfrac{1}{2}\eta_{ac}\eta_{bd} M^{ab}J^{cd}[z]
\end{equation}
where, just as in \eqref{eq:NaiveCharges},
\begin{subequations}
\label{eq:NaiveChargesRep}
\begin{alignat}{1}
\label{eq:NaiveChargesRep-a}
P^a&\,=\,\int_S T^a_b\,u^b\,\measure\mu\,, \\
\label{eq:NaiveChargesRep-b}
J^{ab}[z]&\,=\,\int_S \bigl[(x^a-z^a)T^b_c-(x^b-z^b)T^a_c\bigr]u^c\,\measure\mu\,.
\end{alignat}
\end{subequations}
Here $u$ is the unit timelike normal to $S$ and $\measure\mu=\star u^\flat$ 
(the Hodge dual of the one-form $u^\flat:=\etadown(u)$) is the induced measure 
(3-form) on $S$. Note that only the $J$'s depend on $z$ because only they 
refer to the non-natural (i.e. $o$-dependent) embedding of the Lorentz 
group into the Poincar\'e group. In contrast, the translation group 
$\Trans(\M)$ is normal and hence has a natural place in  the 
Poincar\'e group. Correspondingly, the linear momenta $P^a$ are 
natural and do not depend on any arbitrary choices. Note that 
it immediately follows from \eqref{eq:NaiveChargesRep-b} that 
\begin{equation}
\label{eq:CoadjointRepUnderTrans}
J[z+a]=J[z]-a\wedge P
\end{equation}
which is just the co-adjoint representation of translations
stated in \eqref{eq:AdjointCoAdjointRepsTrans-b}.

\begin{remark}
\label{rem:meaningOfIntegral}
The discussion up to this point answers all the questions  
posed initially in connection with \eqref{eq:NaiveChargesRep}
in the case of Special Relativity. Globally conserved quantities 
(charges) in connection with Poincar\'e symmetry are valued 
in the vector space dual to the Lie algebra and transform 
according to the co-adjoint representation under Poincar\'e 
transformations of the fields to which these quantities belong. 
The splitting of the space in which the charges take their values 
into a ``translational part'' and a ``homogeneous part'' 
is not natural as far as the latter is concerned. Therefore the 
charges of the homogeneous (Lorentz-) part has an additional 
dependence on a spacetime point whose choice fixes the embedding 
of the Lorentz group into the Poincar\'e group. The very notion
of, say, angular momentum depends on the choice of this point.  
\end{remark}  

\newpage
\section{Supplementary conditions and mass centres}
\label{sec:SupplementaryCondition}
The $z$ dependence of $J$ may be used to put further more or less 
physically motivated conditions on $J[z]$ to restrict the choices 
of $z$. Conditions of that sort are known as 
\emph{supplementary conditions} whose aim is to narrow down the 
choices of $z$ to a one-parameter family $z(\lambda)$ which
is timelike and somehow interpreted as the worldline of the body.
This line has many names depending on what supplementary conditions 
one uses. It can be ``centre-of-mass'', ``centre-of-inertia'',
``centre-of-gravity'', ``centre-of-spin'', ``centre-of-motion'',
``centroid'', etc. Early discussions of some of these concepts 
in Special Relativity were given in \cite{Fock:STG} and 
\cite{Born.Fuchs:1940}. For comprehensive discussions 
see \cite{Pryce:2003} and in particular \cite{Fleming:1965}. 

If $u\in{\vecM}_1:=\{v\in V \mid \eta(u,u)=1\}$ is a unit timelike vector 
characterising an inertial frame of reference, we may, e.g., consider 
the supplementary condition (recall that a dot indicates a 
contraction using the Minkowski metric) 
\begin{equation}
\label{eq:SupplementaryConditionFrame-1}
J[z+a]\cdot u
=0\Leftrightarrow J[z]\cdot u-(P\cdot u)\,a+(a\cdot u)\,P=0\,.
\end{equation}
This is equivalent to a linear inhomogeneous equation for $a$
\begin{subequations}
\label{eq:SupplementaryConditionFrame-2}
\begin{equation}
\label{eq:SupplementaryConditionFrame-2a}
\Pi(a)=\frac{J[z]\cdot u}{P\cdot u}\,,
\end{equation}
where
\begin{equation}
\label{eq:SupplementaryConditionFrame-2b}
\Pi=\id-\frac{P\otimes u^\flat}{P\cdot u}
\end{equation}
\end{subequations}
is the projector onto $u^\perp:=\{v\in V \mid v\cdot u=0\}$ 
parallel to $P$ (caution: not parallel to $u$). Hence the 
solution space is one-dimensional timelike line in $\vecM$ 
parallel to $P$: 
\begin{equation}
\label{eq:WorldlineSolutionVector}
a(z,u;\lambda)=
\frac{J[z]\cdot u}{P\cdot u}+\lambda P\,,\qquad \lambda\in\reals \,.
\end{equation}
Its dependence on $z$ immediately follows from  
\eqref{eq:WorldlineSolutionVector} and \eqref{eq:CoadjointRepUnderTrans}:
\begin{equation}
\label{eq:a-DepOn-z}
a(z+b,u;\lambda)=a\bigl(z,u;\lambda+(u\cdot b)/(u\cdot P)\bigr)-b\,.
\end{equation}
Equation \eqref{eq:WorldlineSolutionVector} is a timelike line 
in $\vecM$ that represents the worldline of the 
\emph{centre-of-mass} in $M$ relative to the origin $z$. 
The wordline in $M$ clearly does not depend on $z$ (up to 
reparametrisation) and is simply given by
\begin{equation}
\label{eq:WorldlineSolutionAffine}
\gamma(u;\lambda)=z+a(z,u;\lambda)\,.
\end{equation}
\begin{definition}
\label{def:CentreOfMassWordline}
The curve $\lambda\mapsto\gamma(u;\lambda)$ is called the 
\emph{centre-of-mass} wordline relative to the inertial 
observer $u$.
\qed
\end{definition}

The body's angular momentum with respect to this 
centre-of-mass is  
\begin{equation}
\label{eq:DefSpinRel-u}
S(u):=J[\gamma(u;\lambda)]:=J[z+a(z,u;\lambda)]\,.
\end{equation}
The right-hand side clearly does not depend on $\lambda$ 
since shifting $\lambda$ moves $a(z,u;\lambda)$ in the 
direction of $P$ according to \eqref{eq:WorldlineSolutionVector} 
and hence leaves $J$ unchanged according to 
\eqref{eq:CoadjointRepUnderTrans}. It then also follows 
immediately from \eqref{eq:a-DepOn-z} that the right-hand 
side of \eqref{eq:DefSpinRel-u} does not depend on $z$. 
Hence, as indicated, $S$ only depends on $u$. 

\begin{definition}
\label{def:SpinRelativeToInertialObserver}
$S(u)$ is called the body's \emph{spin} with respect 
to the inertial observer $u$. 
\qed
\end{definition}
Except for its dependence on $u$, this definition meets 
standard Newtonian intuition. Indeed, according to this 
intuition we would call 
\begin{equation}
\label{eq:DefAngMomRel-u}
L(z,u):=a(z,u;\lambda)\wedge P
\end{equation}
the orbital angular momentum relative to $z$ and $u$ 
(there is again no $\lambda$-dependence due to $P\wedge P=0$). 
Equation \eqref{eq:CoadjointRepUnderTrans} then just 
tells us that the total angular momentum is the sum of 
the spin and orbital parts: 
\begin{equation}
\label{eq:AngMomIsSum}
J[z]=L[z+a(z,u;\lambda)]+a(z,u;\lambda)\wedge P=S(u)+L(z,u)\,.
\end{equation}
As in Newtonian mechanics, the $z$-dependence of angular 
momentum resides exclusively in the orbital part. 
But $S$ and $L$ each also depend on $u$, though in such 
a way that their sum is independent of $u$. This gives 
rise to the following 
\begin{remark}
\label{rem:SplittingAngularMomenta}
Unlike in Newtonian Mechanics, the splitting of the total 
angular momentum into a spin ($z$-independent) and an 
orbital ($z$-dependent) part depends on the inertial frame, 
here represented by $u$.
\qed
\end{remark}
Finally, using the expression \eqref{eq:WorldlineSolutionVector} 
for $a$, we get the following expression for the spin part,
\begin{subequations}
\label{def:SpinRelativeInertialObs}
\begin{equation}
\label{def:SpinRelativeInertialObs-a}
S(u):=J[z]-a(z,u;\lambda)\wedge P
=u\cdot\left(P\wedge\frac{J[z]}{P\cdot u}\right)\,,
\end{equation}
which explicitly displays its $u$-dependence. 
Again note that the $z$-dependence of $J$ (given by 
\eqref{eq:SupplementaryConditionFrame-1}) drops out 
due to the wedge product with $P$. The expression on 
the right-hand side of \eqref{def:SpinRelativeInertialObs-a} 
has a simple geometric interpretation, namely that 
of the (tensor-factor wise) projection of $J[Z]$ parallel 
to $P$ onto $u^\perp$, we may also write
\begin{equation}
\label{def:SpinRelativeInertialObs-b}
S(u)=\Pi\otimes\Pi\,\bigl(J[z]\bigr)\,,
\end{equation}
\end{subequations}
where $\Pi$ is as in \eqref{eq:SupplementaryConditionFrame-2a}.
Note that application of $\Pi\otimes\Pi$ cancels the $z$-dependence 
of $J$ and, in exchange, introduces a $u$-dependence. From both 
expressions \eqref{def:SpinRelativeInertialObs} the defining 
equation \eqref{eq:SupplementaryConditionFrame-1} for the 
centre-of-mass, 
\begin{equation}
\label{eq:CentroidDefiningEq}
S(u)\cdot u=-u\cdot S(u)=0\,
\end{equation}
follows trivially.
In \eqref{eq:a-DepOn-z} we already stated the 
obvious dependence of the line $\lambda\mapsto a(z,u;\lambda)$ 
in $\vecM$ on $z$ (which is just like in Newtonian physics).
More interesting, and purely special-relativistic in nature,  
is its dependence on $u$. It is clear from 
\eqref{eq:WorldlineSolutionVector} that any normal timelike vector 
$u\in\vecM_1$ in equation \eqref{eq:WorldlineSolutionVector} 
yields a worldline $\lambda\mapsto \gamma(u;\lambda)$ in $\M$ 
parallel to $P$. As $u$ varies over the 3-dimensional hyperbola $\vecM_1\subset\vecM$ we obtain a bundle of straight lines 
(geodesics) in $\M$ parallel to $P$:
\begin{equation}
\label{eq:BundleOfGeodesics}
\mathcal{B}=
\bigcup_{u\in\vecM_1}\bigcup_{\lambda\in\reals}\ \bigl\{\gamma(u;\lambda)\bigr\}\,.
\end{equation}
In that bundle a particular line $\gamma=\gamma_*$ is 
distinguished, namely that for which $u\propto P$, i.e. 
\begin{equation}
\label{eq:DefRestSystem}
u=u_*:=P/\Vert P\Vert\,.
\end{equation}
Here we use the notation $\Vert P\Vert:=\sqrt{\vert P\cdot P\vert}$.
This is the only timelike direction the body  
determines by itself.\footnote{Here we assume that $P$
is timelike, which essentially means that we assume the 
energy-momentum tensor to satisfy the condition of 
energy dominance.} 
\begin{definition}
\label{def:RestFrame}
The inertial frame for which $u\propto P$ is called the 
body's \emph{rest frame} and 
\begin{equation}
\label{eq:DefRestMass}
M_0:=(u_*\cdot P)/c
\end{equation}
the body's \emph{rest mass}.
The line $\lambda\mapsto\gamma(u_*;\lambda)$, i.e. the 
centre-of-mass in the body's rest frame, is called its 
\emph{centroid}, or wordline of the \emph{centre-of-inertia}~\cite{Fleming:1965}. 
\qed
\end{definition}

Using \eqref{def:SpinRelativeInertialObs} we can immediately 
write down the body's spin relative to its rest frame,
\begin{equation}
\label{eq:DefSpinInRestFrame}
S_*:=u_*\cdot\bigl(J[z]\wedge u_*\bigr)\,,
\end{equation}
which is clearly independent of $z$. With respect to the 
body's centroid, the bundle \eqref{eq:BundleOfGeodesics} of 
wordlines  of mass-centres has a simple geometric 
description:
\begin{theorem}
\label{thm:MoellerDisc}
The intersection of the bundle $\mathcal{B}$
with the hyperplane   
\begin{equation}
\label{eq:Hyperplane}
\Sigma(u_*,\sigma):=\{x\in\M\mid (x-z)\cdot u_*=\sigma\}
\end{equation}
is a 2-disc in perpendicular to the axis of rotation and with radius 
radius is
\begin{equation}
\label{eq:MoellerRadius}
R_M
=\frac{\Vert S_*\Vert}{\Vert P\Vert}
=\frac{\Vert S_*\Vert}{M_0c}\,.
\end{equation}
\end{theorem}
\begin{definition}
\label{def:MoellerRadius}
The radius \eqref{eq:MoellerRadius} is called the 
\emph{M{\o}ller radius}, first defined in~\cite{Moeller:1957}
and also discussed in, e.g., \cite{Dixon:SpecialRelativity}
and \cite{Sexl.Urbantke:RGP}.
It measures the degree to which different inertial 
observers disagree on the spatial location of the 
centre-of-mass perpendicular to the axis of rotation. 
Typical orders of magnitude for M{\o}ller radii will be 
given below.
\qed
\end{definition}

\noindent
\textbf{Proof of Theorem\,\ref{thm:MoellerDisc}}:
Note first that$\Sigma(u_*,\sigma)$ is the hyperplane 
with normal $u_*\propto P$ and timelike distance 
$\sigma$ from the point $z$. As we may choose any 
convenient $z$, we take it to lie on the centroid.
The hyperplane through $z$ is then 
\begin{equation}
\label{eq:MoellerRadiusProof-1}
\Sigma(u_*,\sigma=0)=\{x\in\M\mid (x-z)\cdot u_*=0\}\,.
\end{equation}
Relative to that choice of $z$ (on the centroid) all other 
mass centres have worldlines   
\begin{equation}
\label{eq:MoellerRadiusProof-2}
\gamma(u;\lambda):=
z+\frac{S_*\cdot u}{P\cdot u}+\lambda P\,,
\end{equation}
with $\lambda$ parametrising the individual worldline and 
$u\in\vecM_1$ the different mass-centres. Since $P\cdot S_*=0$ 
the second and third term on the right-hand side are 
perpendicular, so that the wordline $\gamma(u;\lambda)$ 
intersects $\Sigma(u_*,\sigma=0)$ at $\lambda=0$. Hence
\begin{equation}
\label{eq:MoellerRadiusProof-3}
\mathcal{B}\cap\Sigma(u_*,\sigma=0)
=\left\{z+\frac{S_*\cdot u}{P\cdot u}\ \Big\vert\ u\in\vecM_1\right\}
\end{equation}
The claim is that this is a 2-dimensional disc of radius \eqref{eq:MoellerRadius} centred at $z$ which lies in the 
plane perpendicular to the axis of rotation. To see this, we 
parametrise $u$ by its boost-parameters relative to $u_*$, 
i.e., by its rapidity $\rho\in[0,\infty)$ and spatial direction 
$n\in u_*^\perp$, $n^2=1$, so that  
\begin{equation}
\label{eq:MoellerRadiusProof-4}
u=\cosh(\rho)\,u_*+\sinh(\rho)\,n\,.
\end{equation}
Then, assuming $\Vert S_*\Vert\ne 0$,  
\begin{equation}
\label{eq:MoellerRadiusProof-5}
\frac{S_*\cdot u}{P\cdot u}=
\frac{\Vert S_*\Vert}{\Vert P\Vert}\
\frac{S_*\cdot n}{\Vert S_*\Vert}\
\tanh(\rho)\,.
\end{equation}
Note that $n\mapsto\frac{S_*\cdot n}{\Vert S_*\Vert}$ maps 
$u_*^\perp$ into itself. Since it is a non-zero antisymmetric 
endomorphism of the 3-dimensional vector space $u_*^\perp$
it necessarily has a one-dimensional kernel, which is the 
rotation axis (the common fixed-point set of the rotations 
generated by the Lie-algebra element $S_*$) and maps the plane 
perpendicular to that axis into itself. In fact, since we 
divided by $\Vert S_*\Vert$, the map in the plane perpendicular 
to the rotation axis is a rotation by $\pi/2$. Hence, as $n$ runs 
over the unit 2-sphere in $u_*^\perp$ and $\tanh(\rho)$ over the 
intervall $[0,1)$, the image of the map 
$u\mapsto\frac{S_*\cdot u}{P\cdot u}$ becomes the unit 2-disc 
in $u_*^\perp$. \qed
\begin{remark}
\label{rem:SpinTensorVersusVector}
The condition $u_*\cdot S_*=0$ makes $S_*$ effectively a tensor in the 
antisymmetric tensor product of the 3-dimensional space $u_*^\perp$. 
Since $u_*^\perp$ as well as its antisymmetric tensor product are 
3-dimensional, there exists an isomorphism relating them. A perferred one 
is that of the 3-dimensional Hodge duality map, $\tilde\star$, which 
is obtained from the full (4-dimensional) Hodge duality map, denoted 
by $\star$, by first applying $\star$ followed by left contraction 
with $u_*$, i.e., $\tilde\star T:=u_*\cdot\star T=\star(T\wedge u_*)$; 
compare \eqref{eq:InsertionAndHodge} of Appendix\,\ref{sec:Appendix-ExtProdHodge}. 
In this way we can uniquely associate a spin vector 
$\vec S_*$ with the spin-tensor $S_*$ as follows:
\begin{subequations}
\label{eq:SpinVector-1}
\begin{alignat}{3}
\label{eq:SpinVector-1a}
&\vec S_*:&&=-u_*\cdot\star S_*&&=-\star(S_*\wedge u_*)\,,\\
\label{eq:SpinVector-1b}
&S_*&&=-u_*\cdot\star\vec S_*&&=-\star(\vec S_*\wedge u_*)\,.
\end{alignat}
\end{subequations}
Equation \eqref{eq:SpinVector-1a} can be seen as definition of 
$\vec S_*$ and \eqref{eq:SpinVector-1b} as its inverse 
relation. The latter can be obtained from taking the 
$\star$  of the first and using the fact that $\star\circ\star$ 
is the identity on antisymmetric tensors of odd degree in 
even dimensions and Lorentzian signature, which follows from 
combining formulae \eqref{eq:StarSquared} and \eqref{eq:VolumeFormSquared}
of Appendix\,\ref{sec:Appendix-ExtProdHodge}. This gives
\begin{equation}
\label{eq:SpinVector-2}
\star\vec S_*=-S_*\wedge u_*\,.
\end{equation}
Subsequent contraction with $u_*$, using $u_*\cdot S_*=0$, yields \eqref{eq:SpinVector-1b}. In passing we also note that 
the component versions of \eqref{eq:SpinVector-1} are  
\begin{subequations}
\label{eq:SpinVector-3}
\begin{alignat}{2}
&\vec S_*^n && 
=-\tfrac{1}{2}\varepsilon_{abcd}\eta^{dn}S_*^{ab}u_*^{c}\,,\\
&S_*^{mn}&&=-\varepsilon_{abcd}\eta^{cm}\eta^{dn}\vec S_*^au_*^b\,.
\end{alignat}
\end{subequations}
We note from \eqref{eq:SpinVector-1b} that 
\begin{equation}
\label{eq:SpinVector-4}
\vec S_*\cdot S_*=\star(\vec S_*\wedge\vec S_*\wedge u_*)=0\,,
\end{equation}
which means that $\vec S_*$ lies in the intersection of $u_*^\perp$ 
with the kernel of $S_*$. In other words, $\vec S_*$ points along 
the axis of rotation. Finally we note that 
\begin{equation}
\label{eq:SpinVector-5}
\eta(\vec S_*,\vec S_*)=\eta_{ab}\vec S_*^a\vec S_*^b
=-\tfrac{1}{2}\eta_{ac}\eta_{bd}S^{ab}S^{cd}=-\tfrac{1}{2}
\eta\otimes\eta(S_*,S_*)\,.
\end{equation}
By the definition of the normalised inner product on 
antisymmetric tensors (i.e. dividing by $1/p!$ the 
$p$-fold tensor products of $\eta$ on antisymmetric 
$p$-tensors) and setting
\begin{equation}
\label{eq:DefNormForms}
\Vert S_*\Vert:=\sqrt{\vert\langle S_*,S_*\rangle_{\rm norm}\vert}
\end{equation}
we have (recall 
$\Vert\vec S_*\Vert:=\sqrt{\eta(\vec S_*,\vec S_*)}$)
\begin{equation}
\label{eq:SpeinVector}
\Vert S_*\Vert=\Vert\vec S_*\Vert\,.
\end{equation}
This justifies calling $\vec S_*$ the Spin \emph{vector},
which is associated to the (Lie-algebra valued) spin tensor 
$S_*$. 
\qed
\end{remark}

We end this section by justifying the the terminology 
\emph{centre-of-mass}. For this we recall that given
an energy-momentum tensor $\EMT$ and a unit timelike 
direction $u$, then $\EMT(u,u)$ is the spatial 
energy-density in the rest frame of the inertial observer 
represented by $u$. More precisely, let us foliate the affine
space $\M$ by affine hyperplanes 
\begin{equation}
\label{eq:AffineHyperplanes}
\Sigma(u,\sigma):=\{x\in\M\mid (x-z)\cdot u=\sigma\}
\end{equation}
for some given $u\in\vecM_1$ and $z\in\M$. Each $\Sigma(u,\sigma)$ 
is a spacelike hyperplane of Einstein-simultaneity in 
the inertial frame characterised by $u$. It is clearly also 
a Cauchy surface in Minkowski space. The 3-form 
representing the spatial energy-density of $\EMT$ on 
$\Sigma(u,\sigma)$ is then 
\begin{equation}
\label{eq:SpatialEnergyDensity3Form}
\mathcal{E}(u,\sigma)=\EMT(u,u)\star u^\flat\big\vert_{\Sigma(u,\sigma)}\,,
\end{equation}
where $\star u^\flat$ is the measure 3-form on $\Sigma(u,\sigma)$ 
(the Hodge dual to the 1-form $ u^\flat:=\etadown(u):=\eta(u,\cdot)$). 
The first moment of this energy distribution with respect 
to $z$ is 
\begin{equation}
\label{eq:FirstMoment}
m(z,u;\sigma):=
\int_{\Sigma(u,\sigma)}(x-z)\mathcal{E}(u,\sigma)\Big/\int_{\Sigma(u,\sigma)}\mathcal{E}(u,\sigma)\,, 
\end{equation}
where we explicitly indicated all dependencies on $z$, $u$, and 
$\sigma$ and separated the latter by a semicolon to emphasise 
the special meaning of $\sigma$ as ``time-parameter'' labelling 
the different leafs of the foliation orthogonal to $u$. 
The dependence on $z$ is rather trivial: 
$m(z+b,u;\sigma)=m(z,u;\sigma)-b$ so that the set of points 
\begin{equation}
\label{eq:MassCentreAtMomentInTime}
\gamma(u;\sigma)=z+m(z,u;\sigma)
\end{equation}
is independent of $z$. Moreover, from \eqref{eq:FirstMoment} it 
is obvious that $(\gamma(u,\sigma)-z)\cdot u=\sigma$ so that  $\gamma(u,\sigma)\in\Sigma(u,\sigma)$. Note that the construction 
of the ``first moment'' refers to the affine structure of $\M$. 
Given that $\EMT$ satisfies the weak energy-condition we have 
$\EMT(u,u)\geq 0$, so that $\gamma(u,\sigma)$ lies in the convex hull of $\supp(\EMT)\cap\Sigma(u,\sigma)$.

% Hier weiter

Now let us calculate the right-hand side of \eqref{eq:FirstMoment}.
The denominator is, in view of \eqref{eq:NaiveChargesRep-a},
\begin{equation}
\label{eq:q-Denominator}
\int_{\Sigma(u,\sigma)}\EMT(u,u)\star u^\flat=u\cdot P
\end{equation}
independent of $\sigma$ because $P$ is independent of the 
Cauchy surface the integral is taken over. The $a$-th component 
of the  numerator can be transformed as follows 
(calling $\star u^\flat=d\mu$ and using component language)   
\begin{equation}
\label{eq:q-Numerator}
\begin{split}
\int_{\Sigma(u,\sigma)}(x-z)^a\mathcal{E}(u;\sigma)
&=\int_{\Sigma(u,\sigma)} (x-z)^a\,T^{bc}u_bu_c\,d\mu\\
&=\int_{\Sigma(u,\sigma)} 2(x-z)^{[a}\,T^{b]c}u_bu_c\,d\mu\\
&+\int_{\Sigma(u,\sigma)} (x-z)^b\,T^{ac}u_bu_c\,d\mu\\
&=J^{ab}[z]\,u_b+\sigma P^a
\end{split}
\end{equation}  
where we used that $(x-z)^au_a=\sigma$ for 
$x\in\Sigma(u,\sigma)$. In total we get 
\begin{equation}
\label{eq:CentreOfMassDerivation}
\gamma(u;\sigma)=z+\frac{J[z]\cdot u}{P\cdot u}+\frac{P}{P\cdot u}\,\sigma\,,
\end{equation}
which, upon using the new parameter $\lambda:=\sigma/(P\cdot u)$, 
just turns into \eqref{eq:WorldlineSolutionAffine}\eqref{eq:WorldlineSolutionVector}. 
This justifies the term ``centre-of-mass'' in  Definition\,\ref{def:CentreOfMassWordline},
where ``mass'' is to be understood as proportional to energy. 
For a system of point particles this means dynamical mass, not 
rest mass.\footnote{This definition of centre-of-mass,
using the first moment of the dynamical-mass distribution, 
corresponds to cases (c) (for arbitrary $u$) and (d) (for $u=u_*$) 
in \cite{Pryce:2003}. See this reference for a brief historical 
account of other definitions, e.g., based on the first moment 
of the rest-mass distribution, and a discussion of their partly
peculiar properties, like moving mass centres in the zero 
momentum frame. There is also the issue of the Poisson 
structure for the coordinates of mass-centres, linear momentum, 
and spin, which for the mass centres based on dynamical 
mass where first discussed in \cite{Born.Fuchs:1940}. 
Again we refer to \cite{Fleming:1965} for a comprehensive 
discussion.} 
We emphasise again that the essential use of the 
affine structure in this construction. In fact, the very notion 
of ``first'', ``second'', etc.  ``moments'' of a distributions
presuppose such a structure.

\section{Typical M{\o}ller radii}
\label{sec:MoellerRadii}
The ambiguity expressed in \eqref{eq:MoellerRadius} only exists 
for bodies with spin. The formula suggest that for elementary 
particles it may well be of the order of magnitude of other 
radii, but that for laboratory-size or astrophysical bodies 
it is likely to be completely negligible. Let us therefore 
compute a few examples.

A spin-1/2 particle has $\Vert S\Vert=\hbar/2$ and thus 
\begin{equation}
\label{eq:MoellerRadiusSpinHalf}
R_M=
\frac{\hbar}{2M_0c}=\frac{1}{4\pi}\frac{h}{M_0c}=\frac{1}{4\pi}\lambda_C
\end{equation}
where $\lambda_C$ is the particle's Compton wavelength.
If the particle is electrically charged it has a classical charge-radius 
$R_{\rm classical}$ determined by 
\begin{equation}
\label{eq:DefClassicalChargeRadius}
\frac{e^2}{8\pi\varepsilon_0 R_{\rm classical}}=M_oc^2\,.
\end{equation}
Hence we have 
\begin{equation}
\label{eq:MoellerRadiusChargedParticle}
R_M=R_{\rm classical}/\alpha\approx 137 R_{\rm classical}
\end{equation}

Lets look at the Proton: Its experimentally determined 
``proton radius'' (CODATA 2010) is
\begin{equation}
\label{eq:ProtonRadius}
R^{\rm (Proton)}_{\rm charge}=0.87\cdot 10^{-15}\,\mathrm{m}\,.
\end{equation}
Its Compton wavelength is 
\begin{equation}
\label{eq:ProtonComptonWL}
\lambda_{\rm Proton}=1.32\cdot 10^{-15}\,\mathrm{m}\,,
\end{equation}
and its  M{\o}ller radius is
\begin{equation}
\label{eq:MoellerRadiusProton}
R_M^{(\rm Proton)}=\frac{\lambda_{\rm Proton}}{4\pi}
=1.05\cdot 10^{-15}\,\mathrm{m}\, \approx\,
\frac{1}{8}\cdot R^{\rm (Proton)}_{\rm charge}\,.
\end{equation}

In comparison, a homogeneous rigid body of mass $M$ and Radius 
$R$, rigidly spinning at angular frequency $\omega$, has spin 
angular-momentum equal to 
\begin{equation}
\label{eq:BodySpin}
S=\frac{2}{5}\,MR^2\,\omega
\end{equation}
Hence the ratio of its M{\o}ller radius to its geometric 
radius is 
\begin{equation}
\label{eq:RatioMoellerGeometricRadii}
\frac{R_M}{R}=\frac{S}{McR}=\frac{2}{5}\left(\frac{R\omega}{c}\right)\,,
\end{equation}
which shows that this ratio is of the order of magnitude 
of the circumferential velocity in units of the velocity 
of light. Applying this to Earth and Moon (somewhat idealised) 
gives 
\begin{subequations}
\label{eq:MoellerRadiiEarthMoon}
\begin{alignat}{1}
\label{eq:MoellerRadiiEarthMoon-a}
R_M^{(\rm Earth)}&\,=\, 4\,\mathrm{m}\,,\\
\label{eq:MoellerRadiiEarthMoon-b}
R_M^{(\rm Moon)}&\,=\, 1.1\,\mathrm{cm}\,.
\end{alignat}
\end{subequations}
Note that \emph{Lunar Laser Ranging} also locates the moon's 
``position'' within accuracy of centimeters. Hence the 
M{\o}ller radius is not as ridiculously small as one might 
have anticipated it to be for astronomical bodies. In fact,
lets take the fast spinning Pulsar \emph{PSR J1748-2446ad},
whose frequency is 716\,Hz corresponding to a period of 
1.4 milliseconds, for which we get $R\omega/c\approx 0.24$. 
Hence the ratio of its M{\o}ller radius to its geometric 
radius is
\begin{equation}
\label{eq:MoellerRadiusPulsar}
\left(\frac{R_M}{R}\right)_{\rm Pulsar}\approx 0.1\,,
\end{equation}
which is the typical ratio of relativistic effects for 
neutron stars.  

\newpage
\noindent
\addcontentsline{toc}{section}{Appendices}
\textbf{\LARGE Appendices}
\bigskip
\appendix
\section{Exterior products and Hodge duality}
\label{sec:Appendix-ExtProdHodge}
Let $V$ be a real $n$-dimensional vector space, $V^*$ its 
dual space and $T^pV^*=V^*\otimes\cdots\otimes V^*$ its 
$p$-fold tensor product.\footnote{We follow standard tradition 
to define \emph{forms}, i.e. the antisymmetric tensor 
product on the dual vector space $V^*$ rather than 
on $V$. Clearly, all constructions that are to follow could 
likewise be made in terms if $V$rather than $V^*$.}
$T^pV^*$ carries a representation $\pi_p$ of $S_p$, the 
symmetric group (permutation group) of $p$ objects, 
given by 
\begin{equation}
\label{eq:SymmetricGroupAction}
\pi_P:S_p\rightarrow\End(T^pV^*),\quad
\pi_p(\sigma)\bigl(\alpha_1\otimes\cdots\otimes \alpha_p\bigr):=
\alpha_{\sigma(1)}\otimes\cdots\otimes \alpha_{\sigma(p)}
\end{equation}
and linear extension to sums of tensor products. 
On  $T^pV^*$ we define the linear operator of 
antisymmetrisation by  
\begin{equation}
\label{eq:DefAlt-Operator}
\Alt_p:=\frac{1}{p!}\sum_{\sigma\in S_p}\mathrm{sign(\sigma)\,\pi_p}\,,
\end{equation} 
where $\mathrm{sign}:S_p\rightarrow\{1,-1\}\cong\integers_2$ is
the sign-homomorphism. This linear operator is idempotent (i.e. 
a projection operator) and its image of $T^pV^*$ under $\Alt_p$ 
is the subspace of totally antisymmetric tensor-products. 
We write
\begin{equation}
\label{eq:DefExtProductSpace}
\pi_p\bigl(T^pV^*\bigr)=:\bigwedge^pV^*\,.
\end{equation} 
Clearly 
\begin{equation}
\label{eq:DimExtProductSpace}
\dim\left(\bigwedge^pV^*\right)=
\begin{cases}
\binom{n}{p} & \mathrm{for}\ p\leq n\,,\\
0&\mathrm{for}\ p>n\,.
\end{cases}
\end{equation} 
We set 
\begin{equation}
\label{eq:DefExtAlgebra}
\bigwedge V^*:=\bigoplus_{p=0}^n\bigwedge^p V^*\,.
\end{equation}
Let $\alpha\in\bigwedge^pV^*$ and $\beta\in\bigwedge^qV^*$, then we define 
their antisymmetric tensor product 
\begin{equation}
\label{eq:DefWedgeProd}
\alpha\wedge\beta:=\tfrac{(p+q)!}{p!q!}\,\mathrm{Alt}_{p+q}(\alpha\otimes\beta)\in\bigwedge^{p+q}V^*\,. 
\end{equation}
One easily sees that 
\begin{equation}
\label{eq:WedgeProdSymm}
\alpha\wedge\beta=(-1)^{pq}\,\beta\wedge\alpha\,.
\end{equation}
Bilinear extension of $\wedge$ to all of $\bigwedge V^*$ 
endows it with the structure of a real $2^n$-dimensional 
associative algebra, the so-called exterior algebra over $V^*$.   
If $\alpha_1,\cdots,\alpha_p$ are in $V^*$, we have  
\begin{equation}
\label{eq:WedgeOfOneForms}
\alpha_1\wedge\cdots\wedge \alpha_p
=\sum_{\sigma\in S_p}\text{sign}(\sigma)\,
\alpha_{\sigma(1)}\otimes\cdots\otimes \alpha_{\sigma(p)}\,,
\end{equation}
as one easily shows from \eqref{eq:DefWedgeProd} 
and \eqref{eq:WedgeProdSymm} using induction. 

If $\{\theta^1,\cdots, \theta^n\}$ is a basis of $V^*$, 
a basis of $\bigwedge^p V^*$ is given by the following 
$\binom{n}{p}$ vectors
\begin{equation}
\label{eq:InducedBasisOnExtAlg}
\{\theta^{a_1}\wedge\cdots\wedge \theta^{a_p}
\mid 1\leq a_1<a_2<\cdots<a_p\leq n\}\,. 
\end{equation}
An expansion of $\alpha\in\bigwedge^p V^*$ in this basis is 
written as follows  
\begin{equation}
\label{eq:WedgeProdExpansion}
\alpha=:\tfrac{1}{p!}\,
\alpha_{a_1\cdots a_p}\,\theta^{a_1}\wedge\cdots\wedge \theta^{a_p}\,,
\end{equation}
using standard summation convention and where the coefficients 
$\alpha_{a_1\cdots a_p}$ are totally antisymmetric in all indices. 
On the level of coefficients, \eqref{eq:DefWedgeProd} reads 
\begin{equation}
\label{eq:WedgeProdCoeff}
(\alpha\wedge\beta)_{a_1\cdots a_{p+q}}=\tfrac{(p+q)!}{p!q!}
\alpha_{[a_1\cdots a_p}\beta_{a_{p+1}\cdots a_{p+q}]}\,,
\end{equation} 
where square brackets denote total antisymmetrisation in all 
indices enclosed: 
\begin{equation}
\label{eq:VollsAntismmIndizes}
\alpha_{[a_1\cdots a_p]}
:=\frac{1}{p!}\sum_{\sigma\in S_p}\mathrm{sign}(\sigma)\
\alpha_{a_{\sigma(1)}\cdots a_{\sigma(p)}}\,.
\end{equation}

Suppose there is an inner product (non-degenerate symmetric 
bilinear form) $\eta$ on $V$ and the associated dual inner 
product $\eta^{-1}$ on $V^*$ 
(compare Remark\,\ref{rem:MetricAndNotation}). The latter 
extends to an inner product on each $T^pV^*$ 
by 
\begin{equation}
\label{eq:DefInnProdTensors}
\bigl\langle
\alpha_1\otimes\cdots\otimes \alpha_p,\beta_1\otimes\cdots\otimes \beta_p
\bigr\rangle:=
\prod_{a=1}^p\eta^{-1}(\alpha_a,\beta_a)\
\end{equation}
and bilinear extension:
\begin{equation}
\label{eq:DefInnProdTensorsLinExt}
\bigl\langle
\alpha_{a_1\cdots a_p}\ \theta^{a_1}\otimes\cdots\otimes\theta^{a_p}\,,\,
\beta_{b_1\cdots b_p}\ \theta^{b_1}\otimes\cdots\otimes\theta^{b_p}\bigr\rangle=
\alpha_{a_1\cdots a_p}\beta^{a_1\cdots a_p}\,.
\end{equation}
In particular, it extends to each subspace $\bigwedge^p V^*\subset T^pV*$. 
We have 
\begin{equation}
\label{eq:InnProdPureWedges}
\bigl\langle
\alpha_1\wedge\cdots\wedge \alpha_p\,,\,\beta_1\wedge\cdots\wedge\beta_p
\bigr\rangle
:=p!\sum_{\sigma\in S_p}\mathrm {sign}(\sigma)\prod_{a=1}^p\eta(\alpha_a,\beta_{\sigma(a)})
\end{equation}
and hence 
\begin{equation}
\label{eq:InnProdGenForms}
\Bigl\langle
\tfrac{1}{p!}\alpha_{a_1\cdots a_p}\theta^{a_1}\wedge\cdots\wedge\theta^{a_p}\,,\,
\tfrac{1}{p!}\beta_{b_1\cdots b_p}\theta^{b_1}\wedge\cdots\wedge\theta^{b_p}
\Bigr\rangle
=\alpha_{a_1\cdots a_p}\beta^{a_1\cdots a_p}\,.
\end{equation}

In the totally antisymmetric case it is more convenient to 
renormalise this product in a $p$-dependent fashion. One 
sets 
\begin{equation}
\label{eq:RenormInnerProduct}
\bigl\langle\cdot\,,\,\cdot\big\rangle_{\rm norm}\big\vert_{\bigwedge^pV^*}
:=\tfrac{1}{p!}\,\bigl\langle\cdot\,,\,\cdot\big\rangle\big\vert_{\bigwedge^pV^*}
\end{equation}
so that 
\begin{equation}
\label{eq:RenormInnProdGenForms}
\Bigl\langle
\tfrac{1}{p!}\alpha_{a_1\cdots a_p}\theta^{a_1}\wedge\cdots\wedge\theta^{a_p}\,,\,
\tfrac{1}{p!}\beta_{b_1\cdots b_p}\theta^{b_1}\wedge\cdots\wedge\theta^{b_p}
\Bigr\rangle_{\rm norm}
=\tfrac{1}{p!}\alpha_{a_1\cdots a_p}\beta^{a_1\cdots a_p}\,.
\end{equation}

Given a choice $o$ of an orientation of $V^*$ (e.g. induced by an 
orientation of $V$), there is a unique top-form 
$\varepsilon\in\bigwedge^nV^*$ (i.e. a \emph{volume form} for 
$V$), associated with the triple $(V^*,\eta^{-1},o)$, given by 
\begin{equation}
\label{eq:TopForm}
\varepsilon:=\theta^1\wedge\cdots\wedge\theta^n\,,
\end{equation}
where $\{\theta^1,\cdots,\theta^n\}$ is any 
$\eta^{-1}$-orthonormal Basis of $V^*$ in the 
orientation class $o$. The \emph{Hodge duality} 
map at level $0\leq p\leq n$ is a linear isomorphism 
\begin{subequations}
\label{eq:DefHodgeDuality}
\begin{equation}
\label{eq:DefHodgeDuality-a}
\star_p:\bigwedge^pV^*\rightarrow\bigwedge^{n-p}V^*\,,
\end{equation}
defined implicitly by 
\begin{equation}
\label{eq:DefHodgeDuality-b}
\alpha\wedge\star_p \beta=\varepsilon\,\langle\alpha\,,\,\beta\rangle_{\rm norm}\,.
\end{equation}
\end{subequations}
This means that the image of $\beta\in\bigwedge^pV^*$ under 
$\star_p$ in $\bigwedge^{n-p}V^*$ is defined by the 
requirement that \eqref{eq:DefHodgeDuality-b} holds 
true for all $\alpha\in\bigwedge^pV^*$. Linearity is immediate 
and uniqueness of $\star_p$ follows from the fact that 
if $\lambda\in\bigwedge^{n-p}V^*$ and $\alpha\wedge\lambda=0$ for all 
$\alpha\in\bigwedge^pV^*$, then $\lambda=0$. To show existence it 
is sufficient to define $\star_p$ on basis vectors. 
Since \eqref{eq:DefHodgeDuality-b} is also linear in 
$\alpha$ it is sufficient to verify \eqref{eq:DefHodgeDuality-b} 
if $\alpha$ runs through all basis vectors. 

From now on we shall follow standard practice and drop the 
subscript $p$ on $\star$, supposing that this will not cause 
confusion.  

Let $\{e_1,\cdots e_n\}$ be a basis of $V$ and 
$\{\theta^1,\cdots ,\theta^n\}$ its dual basis of $V^*$;
i.e. $\theta^a(e_b)=\delta^a_b$. Let further 
$\{\theta_{1},\cdots ,\theta_n\}$ be the basis of $V^*$
given by the image of $\{e_1,\cdots e_n\}$ under 
$\etadown$ (compare Remark\,\ref{rem:MetricAndNotation}),
i.e. $\theta_a=\eta_{ab}\theta^b$. Then, on the basis 
$\{\theta_{a_1}\wedge\cdots\wedge\theta_{a_p}\mid 1\leq a_1<a_2<\cdots <a_p\leq n\}$
of $\bigwedge^pV^*$ the map $\star$ has the simple form   
\begin{equation}
\label{eq:StarMapOnBasis1}
\star(\theta_{b_1}\wedge\cdots\wedge\theta_{b_p})=\tfrac{1}{(n-p)!}
\varepsilon_{b_1\cdots b_p\,a_{p+1}\cdots a_n}\,\theta^{a_{p+1}}\wedge\cdots\wedge\theta^{a_n}\,.
\end{equation}
This is proven by merely checking \eqref{eq:DefHodgeDuality-b}
for $\alpha=\theta^{a_1}\wedge\cdots\wedge\theta^{a_p}$ and 
$\beta=\theta_{b_1}\wedge\cdots\wedge\theta_{b_p}$. Instead of 
\eqref{eq:StarMapOnBasis1} we can write
\begin{alignat}{1}
\label{eq:StarMapOnBasis2}
\star(\theta^{a_1}\wedge\cdots\wedge\theta^{a_p})
=&\tfrac{1}{(n-p)!}\
\eta^{a_1b_1}\cdots \eta^{a_pb_p}\
\varepsilon_{b_1\cdots b_pb_{p+1}\cdots b_n}\,
\theta^{b_{p+1}}\wedge\cdots\wedge\theta^{b_n}\nonumber\\
=&\tfrac{1}{(n-p)!}\
\varepsilon^{a_1\cdots a_p}_{\phantom{a_1\cdots a_p}a_{p+1}\cdots a_n}
\ \theta^{a_{p+1}}\wedge\cdots\wedge\theta^{a_n}\,,
\end{alignat}
which makes explicit the dependence on $\varepsilon$ 
and $\eta$.

If $\alpha=\frac{1}{p!}\alpha_{a_1\cdots a_p}\theta^{a_1}\wedge\cdots\wedge
\theta^{a_p}$, then $\star\alpha=\tfrac{1}{(n-p)!}
(\star\alpha)_{b_1\cdots b_{n-p}}
\theta^{b_1}\wedge\cdots\wedge\theta^{b_{n-p}}$, where  
\begin{equation}
\label{eq:StarInCoordinates}
(\star\alpha)_{b_1\cdots b_{n-p}}=
\tfrac{1}{p!}\,\alpha_{a_1\cdots a_p}
\varepsilon^{a_1\cdots a_p}_{\phantom{a_1\cdots a_p}b_1\cdots b_{n-p}}\,.
\end{equation}
This gives the familiar expression of Hodge duality in component 
language. Note that on component level the first (rather than 
last) $p$ indices are contracted.

Applying $\star$ twice (i.e. actually $\star_{(n-p)}\circ\star_p$)
leads to the following self-map of $\bigwedge^pV^*$:
\begin{equation}
\label{eq:StarSquared}
\begin{split}
\star\bigl(&\star(\theta^{a_1}\wedge\cdots\wedge\theta^{a_p})\bigr)\\
&=\tfrac{1}{p!(n-p)!}
\varepsilon^{a_1\cdots a_p}_{\phantom{a_1\cdots a_p}a_{p+1}\cdots a_n}
\varepsilon^{a_{p+1}\cdots a_n}_{\phantom{a_{p+1}\cdots a_n}b_1\cdots b_p}\,
\theta^{b_1}\wedge\cdots\wedge\theta^{b_p}\\
&=\tfrac{(-1)^{p(n-p)}}{p!(n-p)!}\,
\varepsilon^{a_1\cdots a_pa_{p+1}\cdots a_n}
\varepsilon_{b_1\cdots b_pa_{p+1}\cdots a_n}\,
\theta^{b_1}\wedge\cdots\wedge\theta^{b_p}\\
&=(-1)^{p(n-p)}\,\langle\varepsilon,\varepsilon\rangle_{\rm norm}\
\theta^{a_1}\wedge\cdots\wedge\theta^{a_p}\,.
\end{split}
\end{equation}
Note that 
\begin{equation}
\label{eq:EpsilonSquared}
\langle\varepsilon,\varepsilon\rangle_{\rm norm}
=\tfrac{1}{n!}\eta^{a_1b_1}\cdots\eta^{a_nb_n}
\varepsilon_{a_1\cdots a_n}\varepsilon_{b_1\cdots b_n}
=(\varepsilon_{12\cdots n})^2/\det\{\eta(e_a,e_b)\}\,.
\end{equation}
This formula holds for any volume form $\varepsilon$ 
in the definition \eqref{eq:DefHodgeDuality-b}, 
independent of whether or not it is related to $\eta$. 

Since the right-hand side of \eqref{eq:DefHodgeDuality-b}
is symmetric under the exchange $\alpha\leftrightarrow\beta$,
so must be the left-hand side. Using \eqref{eq:StarSquared} 
we get 
\begin{equation}
\label{eq:StarStarInnProd1}
\begin{split}
\langle\alpha,\beta\rangle_{\rm norm}\,\varepsilon
&=\alpha\wedge\star\beta
=\beta\wedge\star\alpha
=(-1)^{p(n-p)}\star\alpha\wedge\beta\\
&=\langle\varepsilon,\varepsilon\rangle_{\rm norm}^{-1}\,
 \star\alpha\wedge\star\star\beta
=\langle\varepsilon,\varepsilon\rangle_{\rm norm}^{-1}\,
 \langle\star\alpha\,,\,\star\beta\rangle_{\rm norm}\,\varepsilon\,,
\end{split}
\end{equation}
hence
\begin{equation}
\label{eq:StarStarInnProd2}
\langle\star\alpha\,,\,\star\beta\rangle_{\rm norm}
=\langle\varepsilon,\varepsilon\rangle_{\rm norm}
 \langle\alpha,\beta\rangle_{\rm norm}\,.
\end{equation}
From this and \eqref{eq:StarSquared}) it follows for 
$\alpha\in\bigwedge^pV^*$ and $\beta\in\bigwedge^{n-p}V^*$, 
that
\begin{equation}
\label{eq:StarAdjoint}
\langle\alpha,\star\beta\rangle_{\rm norm}
=\langle\varepsilon,\varepsilon\rangle_{\rm norm}^{-1}
\langle\star\alpha\,,\,\star\star\beta\rangle_{\rm norm}     
=\,(-1)^{p(n-p)}\,\langle\star\alpha,\beta\rangle_{\rm norm}\,.
\end{equation}
This shows that the adjoint map of $\star$ relative 
to $\langle\cdot\,,\,\cdot\rangle_{\rm norm}$ is 
$(-1)^{p(n-p)} \,\star$.  

Formulae \eqref{eq:StarSquared}, \eqref{eq:StarStarInnProd1}\eqref{eq:StarStarInnProd2}, 
and \eqref{eq:StarAdjoint} are valid for general $\varepsilon$ in the definition 
\eqref{eq:DefHodgeDuality-b}. If we chose $\varepsilon$ in the way we did,
namely as the unique volume form that assigns unit volume to an oriented 
orthonormal frame, as does \eqref{eq:TopForm}, then we have 
\begin{equation}
\label{eq:VolumeFormSquared}
\langle\varepsilon,\varepsilon\rangle_{\rm norm}=(-1)^{n_-}
\end{equation}
where $n_-$ is the maximal dimension of subspaces 
in $V$ restricted to which $\eta$ is negative definite;
i.e. $\eta$ is of signature $(n_+,n_-)$. Equation 
\eqref{eq:StarStarInnProd2} then shows that $\star$ 
is an isometry for even $n_-$ and an anti-isometry for 
odd $n_-$ (as for Lorentzian $\eta$ in any dimension).

Finally we note the following useful formula: If $v\in V$
let $i_v: T^pV^*\rightarrow T^{p-1}V^*$ the map which inserts 
$v$ into the first tensor factor. It restricts to a 
map $i_v: \bigwedge^pV^*\rightarrow\bigwedge^{p-1}V^*$. 
Then, for any $\alpha\in\bigwedge^pV^*$, we have   
\begin{equation}
\label{eq:InsertionAndHodge}
i_v\star\alpha=\star(\alpha\wedge v^\flat)\,.
\end{equation}
where $v^\flat:=\etadown(v)$ (compare Remark\,\ref{rem:MetricAndNotation}).
It suffices to prove this for basis elements $v=e_a$ of $V$ 
and $\alpha=\theta^{a_1}\wedge\cdots\wedge\theta^{a_p}$ of $\bigwedge^pV^*$, 
which is almost immediate using \eqref{eq:StarMapOnBasis2}.

\section{Group actions on manifolds}
\label{sec:Appendix-GroupsActions}
Let $G$ be a group and $M$ a set. An \emph{action} of $G$ of $M$ 
is a map
\begin{equation}
\label{eq:GroupAction_Def-1}
\Phi: G\times M\rightarrow M
\end{equation}
such that, for all $m\in M$ and $e\in G$ the neutral element, 
\begin{equation}
\label{eq:GroupAction_Def-2}
\Phi(e,m)=m\,,
\end{equation}
and where, in addition, one of the following two conditions 
hold:
\begin{subequations}
\label{eq:GroupAction_Def-3}
\begin{alignat}{2}
\label{eq:GroupAction_Def-3a}
&\Phi\bigl(g,\Phi(h,m)\bigr)
&&\,=\,\Phi(gh,m)\,,\\
\label{eq:GroupAction_Def-3b}
&\Phi\bigl(g,\Phi(h,m)\bigr)
&&\,=\,\Phi(hg,m)\,.
\end{alignat}
\end{subequations}
If \eqref{eq:GroupAction_Def-1}\eqref{eq:GroupAction_Def-2}
and \eqref{eq:GroupAction_Def-3a} hold we speak of a 
\emph{left action}. A \emph{right action} satisfies   
\index{group action!left}\index{group action!right}
\index{action!left, of group}\index{action!right, of group}
\eqref{eq:GroupAction_Def-1}\eqref{eq:GroupAction_Def-2}
and \eqref{eq:GroupAction_Def-3b}. For a left action we also 
write 
\begin{subequations}
\label{eq:GroupActions_DotNotation}
\begin{equation}
\label{eq:GroupActions_DotNotation-a}
\Phi(g,m)=:g\cdot m
\end{equation}
and for a right action 
\begin{equation}
\label{eq:GroupActions_DotNotation-b}
\Phi(g,m)=:m\cdot g\,.
\end{equation}
\end{subequations}
Equations \eqref{eq:GroupAction_Def-3} then simply become
(group multiplication is denoted by juxtaposition without 
a dot) 
\begin{subequations}
\label{eq:GroupAction_AltDef-3}
\begin{alignat}{2}
\label{eq:GroupAction_AltDef-3a}
&g\cdot(h\cdot m)
&&\,=\,(gh)\cdot m\,,\\
\label{eq:GroupAction_AltDef-3b}
&(m\cdot h)\cdot g
&&\,=\,m\cdot(hg)\,.\\
\end{alignat}
\end{subequations}
Holding either of the two arguments of $\Phi$ fixed we obtain 
the families of maps 
\begin{equation}
\label{eq:GroupAction_ArgFixed-1}
\begin{split}
\Phi_g:\ 
M&\rightarrow M\\
m&\mapsto\Phi(g,m)
\end{split}
\end{equation}
for each $g\in G$, or
\begin{equation}
\label{eq:GroupAction_ArgFixed-2}
\begin{split}
\Phi_m:\
G&\rightarrow M\\
g&\mapsto\Phi(g,m)
\end{split}
\end{equation}
for each $m\in M$. Note that \eqref{eq:GroupAction_Def-2} and \eqref{eq:GroupAction_Def-3} imply that 
$\Phi_{g^{-1}}=\bigl(\Phi_g)^{-1}$. Hence each $\Phi_g$ is a 
bijection of $M$.  The set of bijections of $M$ will be denoted 
by $\Bij(M)$. It is naturally a group with group multiplication 
being given by composition of maps and the neutral element being 
given by the identity map.  Conditions \eqref{eq:GroupAction_Def-2} and \eqref{eq:GroupAction_Def-3a} are then equivalent to the statement 
that the map $G\rightarrow\Bij(M)$, given by $g\mapsto\Phi_g$, is 
a group homomorphism. Likewise, \eqref{eq:GroupAction_Def-2} and \eqref{eq:GroupAction_Def-3b} is equivalent to the statement that 
this map is a group anti-homomorphism.  

The following terminology is standard: The set 
$\Stab(m):=\{g\in G \mid \Phi(g,m)=m\}\subset G$ is called the \emph{stabiliser} of $m$. 
It is easily proven to be a normal subgroup of $G$ satisfying 
$\Stab(g\cdot m)=g\bigl(\Stab(m)\bigr)g^{-1}$ for left and 
$\Stab(m\cdot g)=g^{-1}\bigl(\Stab(m)\bigr)g$ for right 
actions. The \emph{orbit} of $G$ through $m\in M$ is the set 
$\Orb(m):=\{\Phi(g,m) \mid g\in G\}=:\Phi(G,m)$ (also written 
$G\cdot m$ for left and $m\cdot G$ for right action). 
It is easy to see that two 
orbits are either disjoint or identical. Hence the orbits 
partition $M$.  A point $m\in M$ is  called a \emph{fixed point} 
of the action $\Phi$ iff $\Stab(m)=G$.  An action $\Phi$ is called 
\emph{effective} iff $\Phi(g,m)=m$ for all $m\in M$ implies $g=e$;
i.e., ``only the group identity moves nothing''. Alternatively, 
we may say that effectiveness is equivalent to the map 
$G\mapsto\Bij(M)$, $g\mapsto\Phi_g$, being injective; i.e., 
$\Phi_g=\id_M$ implies $g=e$. The action $\Phi$ is called \emph{free}
iff $\Phi(g,m)=m$ for some $m\in M$ implies $g=e$; i.e., 
``no $g\ne e$ fixes a point''. This is equivalent to the 
injectivity  of all maps $\Phi_m:G\rightarrow M$, 
$g\mapsto\Phi(g,m)$, which can be expressed by saying that 
all orbits of $G$ in $M$ are faithful images of $G$. 

Here we are interested in \emph{smooth} actions. For this we need to 
assume that $G$ is a Lie group, 
\index{Lie!group}
that $M$ a differentiable manifold, and that the map \eqref{eq:GroupAction_Def-1} is smooth. We denote 
by $\exp: T_eG\rightarrow G$ the exponential map. For each 
$X\in T_eG$ there is a vector field $V^X$ on $M$, given by 
\begin{equation}
\label{eq:GroupAction_FundVecField}
\begin{split}
V^X(m)
&=\frac{d}{dt}\Big\vert_{t=0}\Phi\bigl(\exp(tX),m\bigr)\\
&=\Phi_{m*e}(X)\,.
\end{split}
\end{equation}
Here $\Phi_{m*e}$ denotes the differential of the map $\Phi_{m}$
evaluated at $e\in G$. $V^X$ is also called the \emph{fundamental vector 
field} on $M$ associated to the action $\Phi$ of $G$ and to $X\in T_eG$.
(We will later write $\Lie(G)$ for $T_eG$, after we have discussed 
\emph{which} Lie structure on $T_eG$ we choose.)    

In passing we note that from \eqref{eq:GroupAction_FundVecField}
it already follows that the flow map of $V^X$ is given by 
\begin{equation}
\label{eq:GroupAction_FundVF-FlowMap}
\Fl^{V^X}_t(m)=\Phi(\exp(tX),m)\,.
\end{equation}
This follows from $\exp(sX)\exp(tX)=\exp\bigl((s+t)X\bigr)$
and \eqref{eq:GroupAction_Def-3} (any of them), which imply 
\begin{equation}
\label{eq:GroupAction_FundVF-FlowMapComp}
\Fl_s^{V^X}\circ\Fl_t^{V^X}=\Fl_{s+t}^{V^X}\,.
\end{equation}
on the domain of $M$ where all three maps appearing in \eqref{eq:GroupAction_FundVF-FlowMapComp} are defined. 
Uniqueness of flow maps for vector fields then suffice to 
show that \eqref{eq:GroupAction_FundVF-FlowMap} is indeed 
the flow of $V^X$. 
 
Before we continue with the general case, we have a closer 
look at the special cases where $M=G$ and $\Phi$ is 
either the left translation of $G$ on $G$, $\Phi(g,h)=L_g(h):=gh$, 
or the right translation, $\Phi(g,h)=R_g(h):=hg$. The corresponding 
fundamental vector fields \eqref{eq:GroupAction_FundVecField} 
are denoted by $V^X_R$ and $V^X_L$ respectively:
\begin{subequations}
\label{eq:GroupActions_DefFundVF-OnGroups}
\begin{alignat}{2}
\label{eq:GroupActions_DefFundVF-OnGroups-a}
&V_R^X(h)
&&\,=\,\frac{d}{dt}\Big\vert_{t=0}\Bigl(\exp(tX)\,h\Bigr)\,,\\
\label{eq:GroupActions_DefFundVF-OnGroups-b}
&V_L^X(h)
&&\,=\,\frac{d}{dt}\Big\vert_{t=0}\Bigl(h\,\exp(tX)\Bigr)\,.
\end{alignat}
\end{subequations}
The seemingly paradoxical labeling of $R$ for left and $L$ 
for right translation finds its explanation in the fact 
that $V^X_R$ is right and  $V^X_L$ is left 
invariant, i.e., $R_{g*}V^X_R=V^X_R$ and $L_{g*}V^X_L=V^X_L$.
Recall that the latter two equations are shorthands for 
\begin{subequations}
\label{eq:GroupAction_LRinvariantVF}
\begin{alignat}{2}
\label{eq:GroupAction_LRinvariantVF-a}
& R_{g*h}V^X_R(h)&&\,=\,V^X_R(hg)\,,\\
\label{eq:GroupAction_LRinvariantVF-b}
& L_{g*h}V^X_L(h)&&\,=\,V^X_L(gh)\,.
\end{alignat}
\end{subequations}
The proofs of \eqref{eq:GroupAction_LRinvariantVF-a} 
only uses \eqref{eq:GroupActions_DefFundVF-OnGroups-a} 
and the chain rule: 
\begin{subequations}
\label{eq:GroupAction_LeftRightInvarianceGroupFundVF}
\begin{equation}
\label{eq:GroupAction_LeftRightInvarianceGroupFundVF-a}
\begin{split}
R_{g*h}V^X_R(h)
&=R_{g*h}\frac{d}{dt}\Big\vert_{t=0}\Bigl(\exp(tX)\,h\Bigr)\\
&=\frac{d}{dt}\Big\vert_{t=0}R_g\Bigl(\exp(tX)\,h\Bigr)\\
&=\frac{d}{dt}\Big\vert_{t=0}\Bigl(\exp(tX)\,hg\Bigr)\\
&=V^X_R(hg)\,.
\end{split}
\end{equation}
Similarly, the proof of 
\eqref{eq:GroupAction_LRinvariantVF-b} starts from 
\eqref{eq:GroupActions_DefFundVF-OnGroups-b}: 
\begin{equation}
\label{eq:GroupAction_LeftRightInvarianceGroupFundVF-b}
\begin{split}
L_{g*h}V^X_L(h)
&=L_{g*h}\frac{d}{dt}\Big\vert_{t=0}\Bigl(h\,\exp(tX)\Bigr)\\
&=\frac{d}{dt}\Big\vert_{t=0}L_g\Bigl(h\,\exp(tX)\Bigr)\\
&=\frac{d}{dt}\Big\vert_{t=0}\Bigl(gh\,\exp(tX)\Bigr)\\
&=V^X_L(gh)\,.
\end{split}
\end{equation}
\end{subequations}
In particular, we have
\begin{subequations}
\begin{alignat}{2}
&V^X_R(g)=R_{g*e}V^X_L(e)&&\,=\,R_{g*e}X\,,\\
&V^X_L(g)=L_{g*e}V^X_R(e)&&\,=\,L_{g*e}X\,,
\end{alignat}
\end{subequations}
showing that the vector spaces of right/left invariant vector
fields on $G$ are isomorphic to $T_eG$. Moreover, the vector spaces 
of right/left invariant vector fields on $G$ are Lie algebras, 
\index{Lie!algebra}\index{algebra!Lie}
the Lie product being their ordinary commutator 
(as vector fields). This is true because the operation of commuting
vector fields commutes with push-forward maps of diffeomorphisms: 
$\phi_*[V,W]=[\phi_*V,\phi_*W]$. This implies that the commutator 
of right/left invariant vector fields is again right/left invariant. 
Hence the isomorphisms can be used to turn $T_eG$ into a Lie algebra,
\index{Lie!algebra}\index{algebra!Lie}
identifying it either with the Lie algebra of right- or left-invariant    
vector fields. The standard convention is to choose the latter.  
Hence, for any $X,Y\in\Lie(G)$, one defines 
\begin{equation}
\label{eq:GroupAction-DefLieAlg-a}
[X,Y]:=[V_L^X,V_L^Y](e)\,.
\end{equation}
$T_eG$ endowed with \emph{that} structure is called $\Lie(G)$.
Clearly, this turns $V_L:\Lie(G)\rightarrow\Vect(G)$, $X\mapsto V_L^X$,  
into a Lie homomorphism:
\index{Lie!homomorphism}
\begin{equation}
\label{eq:GroupAction-LeftCommutator}
V_L^{[X,Y]}=[V_L^X,V_L^Y]\,.
\end{equation}
As a consequence, $V_R:\Lie(G)\rightarrow\Vect(G)$, $X\mapsto V_R^X$, 
now turns out to be an \emph{anti} Lie isomorphism, i.e., 
\index{Lie!anti-isomorphism}
to contain an extra minus sign: 
\begin{equation}
\label{eq:GroupAction-RightCommutator}
V_R^{[X,Y]}:=-\,[V_R^X,V_R^Y]\,.
\end{equation}
This can be proven directly but will also follow from the 
more general considerations below.

On $G$ consider the map 
\begin{equation}
\label{eq:GoupAction_DefConjugation}
\begin{split}
C: G\times G & \rightarrow G\\
(h,g) & \mapsto hgh^{-1}\,.
\end{split}
\end{equation} 
For fixed $h$ this map, $C_h:G\rightarrow G$, $g\mapsto C_h(g)=hgh^{-1}$, 
is an automorphism (i.e., self-isomorphism) of $G$. Automorphisms of 
$G$ form a group (multiplication being composition of maps) which we
denote by $\mathrm{Aut}(G)$. It is immediate that the map $C\rightarrow\mathrm{Aut}(G)$, $h\mapsto C_h$, is a 
homomorphism of groups; i.e., 
\begin{subequations}
\label{eq:GroupAction_InnerAut}
\begin{alignat}{1}
\label{eq:GroupAction_InnerAut-a}
C_e&\,=\,\id_G\,,\\
\label{eq:GroupAction_InnerAut-b}
C_h\circ C_k&\,=\,C_{hk}\,.
\end{alignat}
\end{subequations}

Taking the differential at $e\in G$ of $C_h$ we obtain a linear 
self-map of $T_eG$, which we call $\Ad_h$:
\begin{subequations}
\label{eq:GroupAction_AdjointMap}
\begin{equation}
\label{eq:GroupAction_AdjointMap-a}
\Ad_h:=C_{h*e}: T_eG\rightarrow T_eG\,.
\end{equation}
Differentiating both sides of both equations 
\eqref{eq:GroupAction_InnerAut} at $e\in G$, using the 
chain rule together with $C_k(e)=e$ for the second, 
we infer that 
\begin{alignat}{2}
\label{eq:GroupAction_AdjointMap-b}
\Ad_e&\,=\,\id_{T_eG}\,,\\
\label{eq:GroupAction_AdjointMap-c}
\Ad_h\circ\Ad_k&\,=\,\Ad_{hk}\,.
\end{alignat}
\end{subequations}
This implies, firstly, that each linear map 
\eqref{eq:GroupAction_AdjointMap-a} is invertible, i.e. 
an element of the \emph{general linear group} $\GL(T_eG)$ 
of the vector space $T_eG$, and, secondly, that the map
\begin{equation}
\label{eq:GroupAction_AdjointRep}
\begin{split}
\Ad: G&\rightarrow\GL(T_eG)\\
h&\mapsto\Ad_h
\end{split}
\end{equation}
is a group homomorphism. In other words, $\Ad$ is a 
linear representation of $G$ on $T_eG$, called the 
\emph{adjoint representation}.
\index{group representation!adjoint} 

In \eqref{eq:GroupAction_LRinvariantVF} we saw that $V^X_R$ 
and $V^X_L$ are invariant under the action of right and left 
translations respectively (hence their names). But what happens 
if we act on $V^X_R$ with left and on  $V^X_L$ with right 
translations? The answer is obtained from straightforward 
computation. In the first case we get: 
\begin{subequations}
\label{eq:GroupActions_AdEquivRightLeftVF}
\begin{equation}
\label{eq:GroupActions_AdEquivRightLeftVF-a}
\begin{split}
L_{g*h}\bigl(V^X_R(h)\bigr)
&=L_{g*h}\frac{d}{dt}\Big\vert_{t=0}\Bigl(\exp(tX)\,h\Bigr)\\
&=\frac{d}{dt}\Big\vert_{t=0}\Bigl(g\,\exp(tX)\,h\Bigr)\\
&=\frac{d}{dt}\Big\vert_{t=0}\Bigl(C_g\bigl(\exp(tX)\bigr)\,gh\Bigr)\quad\\
&=V_R^{\Ad_g(X)}(gh)\,,
\end{split}
\end{equation}
where we used \eqref{eq:GroupAction_AdjointMap} in the last and 
the definition of $V_R^X$ in the first and last step. Similarly, 
in the second case we have 
\begin{equation}
\label{eq:GroupActions_AdEquivRightLeftVF-b}
\begin{split}
R_{g*h}\bigl(V^X_L(h)\bigr)
&=R_{g*h}\frac{d}{dt}\Big\vert_{t=0}\Bigl(h\,\exp(tX)\,h\Bigr)\\
&=\frac{d}{dt}\Big\vert_{t=0}\Bigl(h\,\exp(tX)\,g\Bigr)\\
&=\frac{d}{dt}\Big\vert_{t=0}\Bigl(hg\,C_{g^{-1}}\bigl(\exp(tX)\bigr)\Bigr)\\
&=V_L^{\Ad_{g^{-1}}(X)}(gh)\,.
\end{split}
\end{equation}
\end{subequations}
Taking the differential of $\Ad$ at $e\in G$ we obtain a linear 
map from $T_eG$ into $\End(T_eG)$, the linear space of 
endomorphisms of $T_eG$ (linear self-maps of $T_eG$).
\begin{equation}
\label{eq:GroupAction_adointRep}
\begin{split}
\ad:=\Ad_{*e}: T_eG&\rightarrow\End(T_eG)\\
X&\mapsto \ad_X\,.
\end{split}
\end{equation}
Now, we have 
\begin{equation}
\label{eq:GroupAction_adjointRepIsComm}
\ad_X(Y)=[X,Y]
\end{equation}
where the right-hand side is defined in 
\eqref{eq:GroupAction-DefLieAlg-a}. The proof of \eqref{eq:GroupAction_adjointRepIsComm} starts from 
the fact that the commutator of two vector fields can be 
expressed in terms of the Lie derivative of the second 
with respect the first vector field in the commutator, 
and the definition of the Lie derivative. We recall from \eqref{eq:GroupAction_FundVF-FlowMap} that the flow of 
the left invariant vector fields is given by right 
translation: $\Fl^{V^X_L}_t(g)=g\,\exp(tX)$. Then we have 
\begin{subequations}
\label{eq:GroupAction_ProofOfLeftRightComm}
\begin{equation}
\label{eq:GroupAction_ProofOfLeftRightComm-a}
\begin{split}
[X,Y]
&=[V^X_L,V^Y_L](e)\\
&=(L_{V^X_L}V^Y_L)(e)\\
&=\frac{d}{dt}\Big\vert_{t=0}\Fl^{V^X_L}_{(-t)*}
\Bigl(V^Y_L(\Fl_t^{V^X_L}(e))\Bigr)\\
&=\frac{d}{dt}\Big\vert_{t=0}\Fl^{V^X_L}_{(-t)*}
\frac{d}{ds}\Big\vert_{s=0}\Fl^{V^Y_L}_{s}\Bigl(\Fl_t^{V^X_L}(e)\Bigr)\\
&=\frac{d}{dt}\Big\vert_{t=0}\frac{d}{ds}\Big\vert_{s=0}
\exp(tX)\exp(sY)\exp(-tX)\\
&=\frac{d}{dt}\Big\vert_{t=0}\Ad_{\exp(tX)}(Y)\\
&=\ad_X(Y)\,.
\end{split}
\end{equation}
A completely analogous consideration, now using 
$\Fl^{V^X_R}_{t}(g)=\exp(tX)\,g$, allows to compute the 
commutator of the right-invariant vector fields 
evaluated at $e\in G$: 
\begin{equation}
\label{eq:GroupAction_ProofOfLeftRightComm-b}
\begin{split}
[V^X_R,V^Y_R](e)
&=(L_{V^X_R}V^Y_R)(e)\\
&=\frac{d}{dt}\Big\vert_{t=0}\Fl^{V^X_R}_{(-t)*}
\Bigl(V^Y_R(\Fl_t^{V^X_R}(e))\Bigr)\\
&=\frac{d}{dt}\Big\vert_{t=0}\Fl^{V^X_R}_{(-t)*}
\frac{d}{ds}\Big\vert_{s=0}\Fl^{V^Y_R}_{s}\Bigl(\Fl_t^{V^X_R}(e)\Bigr)\\
&=\frac{d}{dt}\Big\vert_{t=0}\frac{d}{ds}\Big\vert_{s=0}
\exp(-tX)\exp(sY)\exp(tX)\\
&=\frac{d}{dt}\Big\vert_{t=0}\Ad_{\exp(-tX)}(Y)\\
&=-\ad_X(Y)\\
&=-[X,Y]\,.
\end{split}
\end{equation}
\end{subequations}
Equation \eqref{eq:GroupAction-RightCommutator} now follows 
if we act on both sides of $[V^X_R,V^Y_R](e)=-[X,Y]$
with $R_{g*e}$ and use \eqref{eq:GroupAction_LRinvariantVF-a}. 

We now return to the general case where $M$ is any manifold and 
the vector field $V^X$ is defined by an action $\Phi$ as in 
\eqref{eq:GroupAction_FundVecField} and whose flow map is given by \eqref{eq:GroupAction_FundVF-FlowMap}. Now, given that 
$\Phi$ is a \emph{right action}, we obtain
\begin{subequations}
\label{eq:GroupAction_GenCommRightLeftAction}
\begin{equation}
\label{eq:GroupAction_GenCommRightLeftAction-a}
\begin{split}
&\bigl[V^X,V^Y\bigr](m)\\
&=(L_{V^X}V^Y)(m)\\
&=\frac{d}{dt}\Big\vert_{t=0}\Fl^{V^X}_{(-t)*}
\Bigl(V^Y(\Fl_t^{V^X}(m))\Bigr)\\
&=\frac{d}{dt}\Big\vert_{t=0}\Fl^{V^X}_{(-t)*}
\frac{d}{ds}\Big\vert_{s=0}\Fl^{V^Y}_{s}\Bigl(\Fl_t^{V^X}(m)\Bigr)\\
&=\frac{d}{dt}\Big\vert_{t=0}\frac{d}{ds}\Big\vert_{s=0}
\Phi\bigl(\exp(tX)\exp(sY)\exp(-tX),m\bigr)\\
&=\frac{d}{dt}\Big\vert_{t=0}\Phi_{m*e}\bigl(\Ad_{\exp(tX)}(Y)\bigr)\\
&=V^{\ad_X(Y)}(m)\\
&=V^{[X,Y]}(m)
\end{split}
\end{equation}
where we used \eqref{eq:GroupAction_FundVF-FlowMap} and \eqref{eq:GroupAction_Def-3b} at the fourth and \eqref{eq:GroupAction_adjointRepIsComm} at the 
last equality. 
Similarly, if $\Phi$ is a \emph{left action}, we have 
\begin{equation}
\label{eq:GroupAction_GenCommRightLeftAction-b}
\begin{split}
&\bigl[V^X,V^Y\bigr](m)\\
&=(L_{V^X}V^Y)(m)\\
&=\frac{d}{dt}\Big\vert_{t=0}\Fl^{V^X}_{(-t)*}
\Bigl(V^Y(\Fl_t^{V^X}(m))\Bigr)\\
&=\frac{d}{dt}\Big\vert_{t=0}\Fl^{V^X}_{(-t)*}
\frac{d}{ds}\Big\vert_{s=0}\Fl^{V^Y}_{s}\Bigl(\Fl_t^{V^X}(m)\Bigr)\\
&=\frac{d}{dt}\Big\vert_{t=0}\frac{d}{ds}\Big\vert_{s=0}
\Phi\bigl(\exp(-tX)\exp(sY)\exp(tX),m\bigr)\\
&=\frac{d}{dt}\Big\vert_{t=0}\Phi_{m*e}\bigl(\Ad_{\exp(-tX)}(Y)\bigr)\\
&=-V^{\ad_X(Y)}(m)\\
&=-V^{[X,Y]}(m)
\end{split}
\end{equation}
\end{subequations}
where we used~\eqref{eq:GroupAction_FundVF-FlowMap} and \eqref{eq:GroupAction_Def-3a} at the fourth and again  \eqref{eq:GroupAction_adjointRepIsComm} at the 
last equality.

Finally we derive the analog of 
\eqref{eq:GroupActions_AdEquivRightLeftVF} in the general case. 
This corresponds to computing the push-forward of $V^X$ under 
$\Phi_g$. If $\Phi$ is a left action we will obtain the analog 
of \eqref{eq:GroupActions_AdEquivRightLeftVF-a}, and the analog 
of \eqref{eq:GroupActions_AdEquivRightLeftVF-b}  if $\Phi$ is
a right action. For easier readability we shall also make use of 
the notation \eqref{eq:GroupActions_DotNotation}. For a left 
action we then get 
\begin{subequations}
\label{eq:GroupActions_AdEquivLeftRightaction}
\begin{equation}
\label{eq:GroupActions_AdEquivLeftRightaction-a}
\begin{split}
\Phi_{g*m}\bigl(V^X(m)\bigr)
&=\Phi_{g*m}\frac{d}{dt}\Big\vert_{t=0}\Phi\bigl(\exp(tX),m\bigr)\\
&=\frac{d}{dt}\Big\vert_{t=0}\Phi\bigl(g\,\exp(tX),m\bigr)\\
&=\frac{d}{dt}\Big\vert_{t=0}\Phi\bigr(C_g(\exp(tX)),g\cdot m\bigr)\\
&=\Phi_{(g\cdot m)*e}\frac{d}{dt}\Big\vert_{t=0}C_g\bigl(\exp(tX)\bigr)\\
&=\Phi_{(g\cdot m)*e}\bigl(\Ad_g(X)\bigr)\\
&=V^{\Ad_g(X)}(g\cdot m)\\
&=V^{\Ad_g(X)}\bigl(\Phi(g,m)\bigr)\,.
\end{split}
\end{equation}
Similarly, if $\Phi$ is a right action, 
\begin{equation}
\label{eq:GroupActions_AdEquivLeftRightaction-b}
\begin{split}
\Phi_{g*m}\bigl(V^X(m)\bigr)
&=\Phi_{g*m}\frac{d}{dt}\Big\vert_{t=0}\Phi\bigl(\exp(tX),m\bigr)\\
&=\frac{d}{dt}\Big\vert_{t=0}\Phi\bigl(\exp(tX)\,g,m\bigr)\\
&=\frac{d}{dt}\Big\vert_{t=0}\Phi\bigr(C_{g^{-1}}(\exp(tX)), m\cdot g\bigr)\\
&=\Phi_{(m\cdot g)*e}\frac{d}{dt}\Big\vert_{t=0}C_{g^{-1}}\bigl(\exp(tX)\bigr)\\
&=\Phi_{(m\cdot g)*e}\bigl(\Ad_{g^{-1}}(X)\bigr)\\
&=V^{\Ad_{g^{-1}}(X)}(m\cdot g)\\
&=V^{\Ad_{g^{-1}}(X)}\bigl(\Phi(g,m)\bigr)\,.
\end{split}
\end{equation}
\end{subequations}

\newpage

\addcontentsline{toc}{section}{References}
\bibliographystyle{plain}
\bibliography{RELATIVITY,HIST-PHIL-SCI,MATH,QM} 
\end{document}